# Platforms for integrated nonlinear optics compatible with silicon integrated circuits

## David J. Moss

*School of Electrical and Computer Engineering, RMIT University, Melbourne, Victoria, Australia 3001*

*david.moss@rmit.edu.au*

and

## Roberto Morandotti

*INRS-EMT, 1650  Boulevard Lionel Boulet, Varennes, Québec, Canada, J3X 1S2*

## Summary

**Nonlinear photonic chips are capable of generating and processing signals all-optically with performance far superior to that possible electronically - particularly with respect to speed. Although silicon-on-insulator has been the leading platform for nonlinear optics, its high two-photon absorption at telecommunications wavelengths poses a fundamental limitation that is an intrinsic property of silicon's bandstructure. We review recent progress in new non-silicon CMOS-compatible platforms for nonlinear optics, focusing on Hydex glass and silicon nitride, and briefly discuss the promising new platform of amorphous silicon. These material systems have opened up many new capabilities such as on-chip optical frequency comb generation, ultrafast optical pulse generation and measurement. We highlight their potential future impact as well as the challenges to achieving practical solutions for many key applications.**

# 1. Introduction

All-optical signal generation and processing [1,2] have been highly successful at enabling a vast array of capabilities, such as switching and de-multiplexing of signals at unprecedented speeds [3,4], achieving parametric gain [5] on a chip, Raman lasing [6], wavelength conversion [7], optical logic [8], all-optical regeneration [9,10], radio-frequency (RF) spectrometers operating at THz speeds [11,12], as well as entirely new functions such as ultra-short pulse measurement [13,14] and generation [15] on a chip, optical temporal cloaking [16], and many others. Phase sensitive functions [14, 17], in particular, will likely be critical for telecommunications systems that are already using phase encoding schemes [18, 19]. The ability to produce these devices in integrated form - photonic integrated circuits (PICs) - will reap the greatest dividends in terms of cost, footprint, energy consumption and performance, where demands for greater bandwidth, network flexibility, low energy consumption and cost must all be met.

The quest for high performance platforms for integrated nonlinear optics has naturally focused on materials with extremely high nonlinearity, and silicon-on-insulator (SOI) has led this field for several years [2]. Its high refractive index allows for tight confinement of light within nanowires that, combined with its high Kerr nonlinearity ($n_2$) has yielded extremely high nonlinear parameters of $\gamma = 200,000$ W$^{-1}$ km$^{-1}$ ($\gamma = \omega\, n_2\, /\, c\, A_{eff}$, where $A_{eff}$ is the effective area of the waveguide, $c$ is the speed of light and $\omega$ is the pump frequency).

As a platform for linear photonics, SOI has already gained a foothold as a foundation for the silicon photonic chip industry [20] aimed at replacing the large network interface cards in optical communications networks, with the ability to combine electronics and photonics on the same chip. These first-generation silicon photonic chips typically employ photonics for passive splitters, filters, and multiplexers and offer the substantial benefit of potentially exploiting the enormous global CMOS infrastructure, although not without challenges [21, 22]. What is clear, however, is that any progress made in this direction for these first generation SOI chips will have a direct benefit on future generation all-optical chips, whether in SOI directly or in other CMOS-compatible platforms. Many of the issues that make CMOS compatibility compelling for linear passive devices apply equally well to all-optical-processing chips.

Indeed, silicon is so attractive as a platform for both linear *and* nonlinear photonics that were it not for one single issue, the quest for the ideal all-optical platform might already be solved. In 2004 it was realized [23] that crystalline silicon (the basis of SOI) suffers from high nonlinear absorption due to (indirect) two-photon absorption (TPA) in telecommunications bands at wavelengths less than 2000 nm (Figure 1). Furthermore, the problem is compounded by the fact that TPA generates free carriers which themselves produce significant linear absorption. This was first realized as being a fundamental limitation, at that time in terms of achieving net Raman gain on a chip [23]. Since then, many methods have been developed for reducing the effect of TPA generated free carriers, such as sweeping carriers out by the use of reverse-biased p-i-n junctions [24], reducing carrier lifetime by ion-implantation, and other approaches.

However, this only solves the secondary effect of free carriers – it has no effect on silicon's *intrinsic* nonlinear figure of merit (FOM = $n_2/\beta \lambda$, where $\beta$ is the TPA coefficient and $\lambda$ the wavelength), which is only 0.3 near 1550 nm [25 - 27] (Figure 1). This represents a *fundamental* limitation – an intrinsic property of silicon's bandstructure. The fact that many impressive all-optical demonstrations have been made in silicon despite its low FOM is a testament to how exceptional its linear and nonlinear optical properties are. Nonetheless, the critical impact of silicon's large TPA was illustrated [28, 29] in 2010 by the demonstration of high parametric gain at wavelengths beyond 2 μm, where TPA vanishes. Indeed, it likely that in the mid-infrared wavelength range where silicon is transparent to both one photon and two photon transitions –between 2 and 6 μm – it will undoubtedly remain a highly attractive platform for nonlinear photonics, despite the influence of higher order absorption [30].

For the telecom band, however, the search has continued for the ideal nonlinear platform. Historically, two important platforms have been chalcogenide glasses [31] and AlGaAs [32]. Chalcogenide glasses have achieved very high performance as nonlinear devices, but pose some unique challenges. Realizing nanowires with ultrahigh nonlinearity (> 10,000 $W^{-1}km^{-1}$) has proven elusive due to fabrication challenges, as has been achieving a material reliability on par with semiconductors. AlGaAs was the first platform proposed for nonlinear optics in the telecom band [32] and offers the powerful ability to tune the nonlinearity and FOM by varying the alloy composition. A significant issue for AlGaAs, however, is that nanowires require very challenging fabrication [33] methods. Nonetheless, both platforms offer significant advantages and will undoubtedly play a role in future all-optical photonic chips.

With the motivation of overcoming this fundamental shortcoming of silicon new platforms for nonlinear optics have recently been proposed [34 - 38] that have achieved considerable success and also offer CMOS compatibility. They are based on silicon nitride SiN and high-index doped silica (trade-named Hydex reference patents). Originally developed for linear optics [39 - 41], these platforms are particularly promising due to their low linear loss, relatively large nonlinearities and, most significantly, negligible nonlinear loss at telecommunication wavelengths [42]. In addition, their high quality CMOS-compatible fabrication processes, high material stability, and the ability to engineer dispersion [36] make these platforms highly attractive.

Indeed, within a short period of time significant progress has been made with respect to their nonlinear performance - particularly in the context of optical frequency comb (OFC) generation in microresonators [43 – 50]. Since the demonstration of OFCs in SiN [36] and Hydex [37] in 2010, this field has proliferated. Extremely wide-band frequency combs [42 - 44], sub 100-GHz combs [46], line-by-line arbitrary optical waveform generation [47], ultrashort pulse generation [15, 49], and dual frequency combs [50] have been reported. In addition to OFC generation, optical harmonic generation [51] has been observed. These achievements have not been possible in SOI at telecom wavelengths because of its low FOM.

Here, we review the substantial progress made towards nonlinear optical applications of these new CMOS-compatible platforms as well as the newly emerging promising platform of amorphous silicon. The high performance, reliability, manufacturability of all of these platforms combined with their intrinsic compatibility with electronic-chip manufacturing (CMOS) has raised the prospect of practical platforms for future low-cost nonlinear all-optical PICs.

## 2. Platforms

Silicon nitride ($Si_3N_4$), a CMOS-compatible material well known in the computer chip industry as a dielectric insulator, has been used as a platform for linear integrated optics [39] for some time. However, only recently [34] has it been proposed as a platform for nonlinear optics. Historically, the challenge for SiN optical devices has been to grow low loss layers thicker than 250 nm, due to tensile film stress. Achieving such thick layers is critical for nonlinear optics since both high mode confinement as well as dispersion engineering for phasematching [36] are needed. Thick (>500 nm) low loss SiN layers were recently grown (Figure 2) by plasma-enhanced chemical vapor deposition (PECVD) [34] as well as by low-pressure chemical vapor deposition (LPCVD) [36] (Figure 3). The latter approach employed a thermal cycling process that resulted in very thick (700 nm) films that yielded nanowires with very low propagation loss (0.4 dB/cm).

The first nonlinear optical studies of SiN waveguides were reported in 2008 [34] showing nonlinear shifting of the resonances in 700 nm thick SiN ring resonators with 200 mW CW optical pump power. Time resolved measurements enabled thermal and Kerr contributions to be separated, resulting in an $n_2 \sim 10\times$ that of silica glass, which is consistent with Millers rule [52]. This value has been validated in subsequent reports of nonlinear optics in SiN nanowires and resonators. Figure 1 also shows the group refractive index and linear dispersion $\beta_2$ as a function of wavelength for different nanowire dimensions. This represents the first report of dispersion engineering in SiN waveguides.

Parametric gain in SiN was first demonstrated [36] in low loss nanowires by centering the pump for the FWM process in the anomalous group-velocity dispersion (GVD) regime near the zero-GVD point (Figure 3). This allowed for broad-bandwidth phase matching, and hence signal amplification, over a wide range of wavelengths. Net gain was achieved in long (6 cm) SiN waveguides with a nonlinear $\gamma$ parameter of 1,200 $W^{-1}km^{-1}$ and a zero-GVD point near 1,560 nm. An on/off signal gain as high as 3.6 dB was observed over a 150-nm bandwidth, and since the total propagation loss through the waveguides was 3 dB, this represented net parametric amplification.

Hydex glass was developed [41, patents] as a low loss CMOS compatible optical platform primarily for advanced linear filters. Its refractive index range of $n$ = 1.5 to 1.9 is slightly lower than SiN, being comparable to SiON, and so a buried waveguide geometry is typically used rather than nanowires. Nonetheless, the core-cladding contrast of 17% still allows for relatively tight bend radii of 20 $\mu$m. Its proprietary composition is primarily aimed at reducing the need for annealing by reducing the effect of N-H bonds - the main source of absorption loss in the telecom band. This platform has resulted in extraordinarily low linear propagation losses of 5 - 7 dB/meter, allowing for the use of extremely long waveguide spirals [14, 42, 53]. Figure 4 shows a schematic of a 45-cm-long spiral waveguide contained within a square area of 2.5 mm x 2.5 mm, pigtailed to single-mode fiber via low loss on-chip beam expanders and a SEM picture of its cross section (before cladding deposition). The films were fabricated with CMOS compatible processes that yielded exceptionally low sidewall roughness in the core layer. Self-phase modulation experiments [42] yielded a Kerr nonlinearity of $n_2$ = 1.15x10$^{-19}$ m$^2$/W, or ~ 4.6 $\times$ silica glass, and roughly half as large as that for SiN, with a nonlinearity parameter $\gamma \cong 233W^{-1}km^{-1}$ (~ 200 $\times$ standard single-mode telecommunications fiber). This enhancement in $n_2$ is, like SiN, in agreement with Miller's rule [52], meaning that the proprietary chemistry of Hydex is not relevant to its nonlinear optical performance. The waveguides were engineered to yield anomalous dispersion [54] (Figure 4), critical for efficient, wide bandwidth FWM over most of the C-band with zero-GVD points being 1600 nm for TE polarization and 1560 nm for TM. This resulted in a large FWM wavelength tuning range with efficient parametric gain of +15dB and a signal to idler conversion efficiency of +16.5dB (Figure 4) [53].

Table I summarizes the linear and nonlinear properties of these CMOS compatible platforms. The empirical Millers rule [52] predicts that n$_2$ increases as the 4$^{th}$ power of $n_0$. The price of increasing $n_0$ in order to increase the Kerr nonlinearity is that this also tends to increase propagation losses. The relatively low refractive indices of Hydex (1.75) and $Si_3N_4$ (2.0) result in extremely low linear and nonlinear losses – their greatest strength - but also results in fairly modest values of $n_2$. The low linear and nonlinear losses has largely accounted for the success of these platforms for nonlinear optics in the telecommunications band, resulting in a combination of many factors that include negligible TPA, a moderately high nonlinearity, the ability to engineer dispersion, their high quality growth and fabrication processes, and their excellent material reliability. However, the *ultimate* platform would be one with a nonlinearity ideally larger than silicon *and* with a much larger FOM.

While amorphous silicon has been studied as a nonlinear material for some time [55], and developed as a platform for linear photonics in the telecom band for a number of years [56, 57], only recently has it been proposed [58] as an alternative to SOI for nonlinear optics in the telecom band, with the hope that a-Si could possibly offer lower TPA than silicon. Table II surveys the measured nonlinear properties for a-Si, which show a near universal improvement in both the nonlinearity and FOM over c-Si. Although initial measurements yielded a FOM no better than c-Si (~0.5) [58, 59], more recent results have yielded FOMs from 1 [60] to as high as 2 [61, 62]. This has enabled very high parametric gain of more than +26dB over the C-band [63]. However, to date a key issue for this material has been a lack of stability resulting in a significant degradation in performance over relatively short timescales [64].

Very recently, a-Si nanowires were demonstrated [65] that displayed a combination of high FOM, high nonlinearity, and good material stability at telecom wavelengths. Figure 5 shows a cross section SEM picture of a hydrogenated amorphous silicon nanowire (a-Si:H) waveguide fabricated in a 200mm CMOS pilot line at CEA-LETI, France. The a-Si:H film was deposited by plasma enhanced chemical vapor deposition (PECVD) at 350C on 1.7µm oxide deposited on a bulk wafer. After deposition of a silica hard mask, two steps of 193nm DUV lithography and HBr silicon etching were used to define grating couplers that were well aligned with serpentine waveguides with varying lengths (1.22 cm to 11 cm). The fabricated waveguides are ~ 220nm in thickness and ~ 500nm in width. A 500nm oxide was deposited to provide an upper cladding. The group velocity dispersion for the TE mode confined within a 500nm×220nm nanowire was calculated with FEMSIM, yielding an anomalous second-order dispersion parameter $\beta_2$=-4.2×10-25 s$^2$/m at λ=1550nm.

Figure 5 shows a schematic of the experimental setup used for the measurement of both the linear and nonlinear propagation characteristics of **the** a-Si:H nanowires. A mode-locked fiber laser with near transform limited ~1.8ps long pulses at a repetition rate of 20MHz at 1550nm was coupled into the TE mode of the nanowires via in-plane gratings. The fiber to waveguide coupling loss per coupler was ~ 10.6dB and 12.4dB per entry and exit, respectively - higher than expected due to the grating couplers not being optimized. The propagation loss of the TE mode was measured to be about 4.5 dB/cm, via a cut-back method on serpentine waveguides with lengths varying from 1.22 cm to 11 cm. A comparison with linear measurements performed on 1.3mm long straight nanowires yielded a loss contribution due to the bends (10 μm radius) of about 4dB, i.e. on the order of 0.04dB/ bend.

To determine the nonlinear parameters of the waveguides a series of self- phase modulation (SPM) measurements in a 1.22 cm long nanowire were performed with a coupled peak power up to ~3W. The output spectrum was then measured as a function of input power.

Figure 5 shows the measured output spectra of the pulses for three coupled peak powers as well as the experimental contour plots of the output spectra as a function of coupled peak power ranging between 0.03W and 3W. Strong spectral broadening is observed, which is the signature of self-phase modulation of the pulse propagating along the nonlinear nanowire. At higher coupled powers (>1.8W), the spectral broadening of the pulse was limited by the spectral transfer function of the grating couplers, which had a 3dB bandwidth of around ~35nm centered near 1550nm. Split-step-Fourier method simulations were used to solve the nonlinear Schrödinger equation governing the propagation of the picosecond optical pulse in the nonlinear waveguide, in the presence of second-order dispersion and TPA. The impact of dispersion was negligible, as expected for a $\beta_2$= - 4.2×10-25s$^2$/m, associated with a dispersion length exceeding 7m for the 1.8ps pulses - well over the physical length of the waveguide. Figure 5 shows the output spectra from the simulations, showing good agreement with measurement, when taking the TPA contribution stated above and a nonlinear waveguide parameter $\gamma$ =1200/W/m, associated with a Ker nonlinearity of $n_2 = 2.1$ x 10-17 $m^2$/W. The low TPA of a-Si:H was also reflected in the absence of any blue shift in the output spectra. Figure 5 also shows the output spectra taken over a period of several hours, showing no degradation in the nonlinear characteristics. Experimental measurements of self-phase modulation and nonlinear transmission yielded a record high nonlinear FOM of 5 – over 10 x that of SOI – together with an $n_2$ of 3 to 4 times c-Si.

The ability for a:Si to display both high nonlinearity and FOM has surprised many researchers, since in most materials, higher nonlinearity is accompanied by higher TPA. The key to the "anomalous" behavior of amorphous silicon may reside in the difference between *direct* and *indirect* transitions. In direct bandgap semiconductors (GaAs/AlGaAs [66]), below the bandgap both $n_2$ and TPA <u>*always*</u> increase with decreasing bandgap energy. For these simple direct bandgap semiconductors this is largely a result of Kramers - Kronig relations (adapted for nonlinear response functions [67]) that predict that increasing the bandgap to decrease nonlinear absorption also decreases the nonlinear response. In Si, however, the situation is more complex since TPA in the telecom band arises solely from *indirect* transitions, while $n_2$ (the real part of $\chi^{(3)}$) is dominated by *direct* transitions [67]. The direct bandgap in Si is $> 3.2$ eV while the indirect bandgap is only 1.1eV - well below the two-photon energy (1.6eV) in the telecom band. For a:Si, it is likely that the effective *direct* "bandgap" (or mobility gap, being amorphous) is decreased compared to Si, thus increasing $n_2$, while the contribution of *indirect* transitions is greatly reduced, thus decreasing the TPA. This is entirely plausible since indirect transitions involve phonons which depend on long range order which, even for optical phonons, is clearly reduced in a:Si. Figure 5 shows the band structure of silicon and illustrates the principle of simultaneously increasing the nonlinearity and FOM.

A key goal for all-optical chips is to reduce device footprint and operating power, and the dramatic improvement in the FOM of a-Si raises the possibility of using slow-light structures [68, 69] to allow devices to operate at mW power levels with sub-millimeter lengths.

Finally, we note that the broad use of the description "CMOS compatible" in this context is intended to reflect a general compatibility in terms of growth temperatures (< 400 C) and materials that are familiar and used in the CMOS process (silicon nitride, silicon oxynitride). It does not address the complexities and challenges of integrating optical and electronic devices with substantially different size scales, nor does it address the challenges of adapting CMOS production lines to optical device fabrication, both of which have been discussed at length [20-22]. A central issue in terms of integrating waveguides and nanowires with electronic components is the rather thick nature of both the core and cladding films. In this regard, SiN, with a higher refractive index contrast of about 0.5, offers a significant advantage over Hydex where the index contrast is only about 0.3. Both glasses, however, require noticeably thicker layers (both core and cladding) than SOI and this is probably a key area where SOI out-performs these platforms. Nonetheless, the concept of CMOS compatibility presented here, that these new platforms adhere to, is a powerful one that will go a long way towards enabling the broad application of CMOS techniques and manufacturing infrastructure to nonlinear photonic chips.

## 3. Low Power Nonlinear Optics in Ring Resonators

Optical microcavities including ring resonators can dramatically enhance nonlinear processes [70] and offer a powerful approach to greatly reducing operating powers for many applications such as frequency comb generation for spectroscopy and metrology [70,71]. Microcavities are particularly conducive to enhancing FWM processes because FWM typically involves a pump, signal and idler beams with frequencies ($\omega_{Pump}$, $\omega_{Idler}$, $\omega_{Signal}$) related by energy conservation: $\omega_{Idler} = 2\omega_{Pump} - \omega_{Signal}$. This process can occur with (classical) or without (spontaneous) a separate input signal at $\omega_{Signal}$, the classical effect being much stronger and the basis for most all-optical signal processing. Very low CW operation was achieved in silica and single crystal micro-toroids and spheres with Q-factors from $10^7$ to $10^{10}$ [70, 71]. For microcavities, achieving phase-matching is critical, which is equivalent to achieving equal resonance spacings (constant FSR) (with due allowance for the Kerr induced resonance shifts [36]) that results in pump, signal and idler all being in resonance – a triple resonance that greatly reduces the power requirement for the round-trip parametric gain to exceed the loss, thus producing oscillation. A comprehensive review of the use of microcavities for frequency comb generation is found in [48].

The first demonstration of low power nonlinear optics in these CMOS platforms was in Hydex [38] in 2008, where low power CW nonlinear optics was achieved via FWM in integrated ring resonators. Figure 6 shows a four port micro-ring resonator, radius $\cong$ 48μm, along with a scanning electron microscope (SEM) cross section of the waveguide, having similar dimensions and composition to the spiral structure. Figure 6 also shows the transmission spectrum of the resonator showing a Q-factor of $\cong$ 65,000, a free-spectral range (FSR) of 575 GHz, and a full-width-half-maximum (FWHM) of 3GHz. The bus waveguides, buried in $SiO_2$ beneath the ring, couple light in and out of the resonator.

Figure 7 shows the experimental setup as well as the basic physics behind FWM in ring resonators. The experiments were conducted the ring resonator using only a few mW of CW pump power, with the signal tuned to an adjacent resonance to the pump. FWM was achieved near 1558nm with 5mW of CW pump power (in the waveguide) and a signal power of 570μW tuned to an adjacent resonance. This yielded an idler that was almost exactly on resonance indicating that the dispersion in the system was indeed negligible and confirmed that the process was triply resonantly enhanced.

The experiments were performed with about 5mW [38] of CW pump power tuned to a ring resonance at 1553.38nm (TM polarization) and directed to the INPUT port, while a signal laser (TM polarization) with a power of 550μW was tuned to an adjacent resonance at 1558.02 nm and directed to the ADD port. Figure 7 shows the resulting output power spectra (TM polarization) as recorded from both the THROUGH and the DROP ports after the pigtail; experiments carried out with TE polarized modes led to similar results. Two detectable idlers were generated by the FWM process - the first-idler power in the bus waveguide was $\cong$ 930 pW, whereas the second idler had a power of $\cong$100 pW. The ratio of the powers for the two idlers agrees remarkably well with the ratio of the pump to the signal power, $P_{idler(-1)}/P_{idler(-2)} \cong P_{pump}/P_{signal}$, as expected for FWM. The first idler was almost exactly on resonance at 1548.74 nm, indicating that the dispersion in the system is indeed negligible.

The theoretical conversion efficiency $\eta$ takes into account the cavity enhancement factor due to the ring geometry:

$$\eta \equiv \frac{P_{idler}}{P_{signal}} = \left|2\pi R\gamma\right|^2 P_{pump}^2 \cdot \left(FE_p\right)^4 \cdot \left(FE_s\right)^2 \cdot \left(FE_i\right)^2 \qquad (1)$$

$$FE_\mu = \frac{\sqrt{2(1-\sigma_\mu)}}{2(1-\sigma_\mu)+\alpha_\mu\pi R} \qquad\qquad \sigma_\mu = \left(1-\frac{\pi}{2Finesse_\mu}\right)\exp\left(\frac{\alpha_\mu\pi R}{2}\right) \qquad (2)$$

where $R$ is the ring radius, $FE_\mu$ describes the field enhancement of the ring and $\alpha_\mu$ is the mode linear loss coefficient for the mode $\mu$. Finally, $\sigma_\mu$ is the self-coupling coefficient between ring and channels. The product $(FE_p)^4(FE_s)^2(FE_i)^2$ identifies an overall field enhancement factor.

Figure 7 shows the external conversion efficiency (including coupling losses), along with the external conversion efficiency of comparable experiments carried out in silicon ring resonators [72], estimating the efficiency from the coupling loss. Although the conversion efficiency of the Hydex device was predictably lower than the SOI ring resonators due to the much lower γ, the negligible nonlinear absorption potentially allows the scaling of pump powers to levels at which high bit rate all-optical signal processing in silicon nanowires is typically performed [73,74].

# 4. Microresonator-Based Frequency Combs

The area where these platforms have arguably had the greatest impact is in integrated OPOs based on ring resonators. These devices have significant potential for many applications including spectroscopy and metrology as well as the ability to provide an on-chip link between the RF and optical domains [75]. As discussed above, micro-cavities enhance nonlinear optical processes, such as FWM involving a continuous-wave (CW) pump, signal and idler beams with frequencies ($\omega_{Pump}$, $\omega_{Idler}$, $\omega_{Signal}$) related by energy conservation: $\omega_{Idler} = 2\omega_{Pump} - \omega_{Signal}$ . Achieving phase-matching of the propagation constants for three interacting waves is essential for efficient FWM, which for the microcavities is equivalent to having near-equal resonance spacings, or a constant FSR, with due allowance for the Kerr-induced resonance shifts [36]. This results in the pump, signal and idler waves all being in resonance – a triple resonance that greatly reduces the power requirement for the round-trip parametric gain to exceed the loss, thus producing oscillation. Phase matching, or low and anomalous waveguide dispersion, can be achieved with a suitable design of the waveguide cross-section.

Reports in early 2010 of OPOs in both SiN [36] and Hydex [37] opened the door to achieving these devices in practical CMOS-compatible integration platforms with much lower Q factors than previous micro-cavity oscillators, and hence with much less sensitivity to environmental perturbations and without the need for delicate tapered fibre coupling. This will allow these devices to benefit not just the scientific community but telecommunications, computing, and precision spectroscopy and timekeeping.

Figure 8 shows a four port micro-ring resonator based in Hydex glass with a Q factor of 1.2 million [37] along with a scanning electron microscope (SEM) of the waveguide cross section and the corresponding optical transmission spectrum. Figure 9 shows the experimental setup [37] for demonstrating an on-chip OPO using an externally pumped laser, highlighting the use of soft thermal locking to achieve resonant coupling between the pump laser and the cavity resonances.

The optical frequency comb generated by the Hydex device (Figure 10) exhibited a very wide spacing of almost 60 nm when pumped at 1,544.15 nm, in the anomalous GVD regime (Figure 3). The output power versus pump power shows a very high single line differential slope efficiency above threshold of 7.4%, with a maximum power of 9 mW achieved in all oscillating modes out of both ports, representing a remarkable absolute total conversion efficiency of 9%. When pumping at a slightly different wavelength closer to the zero-GVD wavelength (but still anomalous), the device oscillated with significantly different spacing of 28.15 nm.

Achieving phase-matching of the propagation constants for three interacting waves is essential for efficient FWM, which for the microcavities is equivalent to having near-equal resonance spacings, or a constant FSR, with due allowance for the Kerr-induced resonance shifts. This results in the pump, signal and idler waves all being in resonance – a triple resonance that greatly reduces the power requirement for the round-trip parametric gain to exceed the loss, thus producing oscillation. Phase matching, achieved by obtaining low and anomalous waveguide dispersion can be realised with a suitable design of the waveguide cross-section. These observations are consistent with parametric gain based on a combination of FWM and MI described above, where the spacing depends on the waveguide dispersion characteristics and agrees well with calculations. Oscillation begins via modulational instability (MI) gain – essentially pure spontaneous (degenerate) FWM (with only a pump present). One oscillation is achieved, "cascaded" FWM among different cavity modes takes over, resulting in the generation of a frequency comb of precisely spaced modes in the frequency domain. However, the enhancement of the fields in the cavity (all fields in the case of phase matching) that is responsible for lowering the oscillation threshold, also enhances the nonlinear losses, and it is primarily for this reason that oscillation in silicon ring resonators in the telecom band where the FOM < 1, has not been achieved.

Figure 11 shows the calculated MI gain curve when pumping at 1544nm with 54mW of power, using the experimentally measured dispersion curves. This shows a gain peak at 1590nm – close to the observed initial oscillation peaks. This confirms that oscillation begins via MI gain near the peak. Once oscillation is achieved, "cascaded" FWM can occur among many different cavity modes resulting in the generation of a frequency comb, of precisely spaced modes in frequency.This illustrates the degree of freedom one can achieve in varying the frequency comb spacing – chiefly through dispersion engineering, and so one is not restricted by the FSR of the resonator itself. The trade-off is that MI generated combs can themselves further seed sub-combs that are poorly related in terms of coherence to the original comb, limiting the degree to which modelocking, or ultrashort pulse generation, can be achieved [47].

Optical parametric oscillation was simultaneously reported in SiN [36] by resonantly pumping the rings with CW light near 1550 nm using a soft 'thermal lock' process in which the cavity heating is counteracted by diffusive cooling. Figure 12 shows schematics of 2-port SiN micro-ring resonators: [36] a 58 µm radius resonator (Q factor = 500,000, FSR = 403GHz) with dimensions designed to yield anomalous GVD in the C-band with a zero GVD point at 1,610 nm, and a 40 µm radius resonator with Q factor of 1.3 million, along with the fibre-pigtailed coupling waveguide used to achieve robust and low-loss coupling of light into and out of the devices along with the transmission spectrum for transverse-magnetic (TM) modes. Oscillation of multiple lines over a very broad (>200 nm) wavelength range was achieved (Figure 12) at a pump threshold of 50 mW. Eighty-seven new frequencies were generated between 1,450 and 1,750 nm, corresponding to wavelengths covering the S, C, L and U communications bands. Several designs were employed with different ring radii, or FSR. A smaller ring with a Q factor of 100,000 generated oscillation in 20 resonator modes when pumped with modest input powers (150 mW) with THz mode spacing. These results represent a significant step toward a stabilized, robust integrated frequency comb source that can be scaled to other wavelengths.

# 5. Advanced Frequency Comb Generation

Since these initial demonstrations of multiple wavelength oscillation, significant progress has been made in advanced comb generation, including both very wide bandwidth octave spanning combs [45] and very low (sub 100GHz) FSR spacing combs [46].

The development of microresonator-based frequency combs with a free spectral range (FSR) significantly less than 100 GHz is critical to provide a direct link between the optical and electrical domains in order to produce highly stable microwave signals detectable with photodiodes. The challenge is that simple ring geometries with sub-100GHz FSR spacings do not fit on typical single e-beam fields and so novel ring geometries such as spirals need to be employed. Figure 13 shows spiral ring resonators with unique geometries for different FSRs below 100GHz [46], all having a constant semicircular coupling region to enable critical coupling between the bus and resonator, independent of path length. Bends in the resonators had radii > 100 μm to ensure that bend induced dispersion was negligible, a critical requirement for proper operation of the frequency comb. The experimental spectra for 80, 40, and 20 GHz combs are shown in Figure 13, typically requiring about 2W pump power to fill the entire comb spans.

Octave-spanning frequency combs are of great interest for spectroscopy, precision frequency metrology, and optical clocks and are highly desirable for comb self-stabilization using $f$ to $2f$ interferometry for precision measurement of absolute optical frequencies [76]. Figure 14 shows an optical frequency comb in a SiN ring resonator spanning more than an octave [45] from 1170 to 2350 nm, corresponding to 128 THz, achieved by suitable dispersion engineering and employing higher pump powers of up to 400 mW inside the waveguide, detuned slightly from a cavity resonance. Figure 14 shows the simulated dispersion for nanowires with varying widths (1200, 1650, and 2000 nm) indicating that large anomalous-GVD bandwidths spanning nearly an octave are possible with appropriate design. These results represent a significant step toward a stabilized, robust integrated frequency comb source that can be scaled to other wavelengths.

# 6. Supercontinuum Generation

By injecting ultrashort modelocked pulse trains into suitably designed waveguides very broadband spectra can be generated. This super-continuum (SCG) spectra also can result in octave-spanning frequency combs and has been of great interest for spectroscopy, precision frequency metrology, and optical clocks and is highly desirable for comb self-stabilization using $f$ to $2f$ interferometry for precision measurement of absolute optical frequencies [76 - 80]. Wide bandwidth SCG has been demonstrated in microstructured fibers [81-83], ChG waveguides [84], periodically poled lithium niobate (PPLN) [85], and in Si [86-88]. Hydex and SiN offer advantages of much lower linear and nonlinear loss as well as transparency well into the visible.

Figure 15 compares the broadband frequency combs with SCG in Hydex waveguides [89] where a spectral width > 350 nm was achieved (limited by experimental measurement capability), with that generated in 1100-nm wide SiN nanowires [90] that yielded a spectrum spanning 1.6 octaves, from 665 nm to 2025 nm, pumping at 1335 nm. The SiN results in particular represent the broadest recorded SCG to date in a CMOS compatible chip. Both of these results were enabled by a high effective nonlinearity, negligible TPA and most significantly very flexible dispersion engineering. SCG is significantly enhanced if the pulse is launched near a zero group-velocity dispersion (GVD) point or in the anomalous GVD regime. The former minimizes temporal pulse broadening, thereby preserving high peak powers and thus maintaining a strong nonlinear interaction. The latter regime enables soliton propagation, whose dynamics can contribute to spectral broadening. The higher flexibility of dispersion engineering in SiN as compared to Hydex is partly enabled by its higher available core/cladding index contrast and is probably one of its most significant advantages.

# 7. Comb coherence and dynamic properties

As with conventional modelocked lasers, parametric frequency combs can also potentially serve as sources of ultrashort laser pulses that, depending on the pump laser and material system, can produce ultrashort pulses from the visible to the mid-infrared at repetition rates in the GHz to THz regimes. The last few years have seen significant breakthroughs in understanding the dynamics and coherence behavior of frequency comb formation [44, 47, 48]. These reports have revealed complex and distinct paths to comb formation that can result in widely varying degrees of coherence, wavelength spacing, and RF stability of these sources. This field has recently been highlighted by the achievement of an ultrastable, ultrashort optical pulse source via modelocking [15] based on an integrated microcavity. Understanding and harnessing the coherence properties of these monolithic frequency comb sources is crucial for exploiting their full potential in the temporal domain, as well as for generally bringing this promising technology to practical fruition.

The first investigation of the coherence properties of these on-chip oscillators [47] focused on phase tuning of the comb via programmable optical pulse shaping (Figure 16). This "line-by-line" pulse shaping of microresonator frequency combs was enabled by the relatively large mode spacings and represented a significant development in the field of optical arbitrary waveform generation [77-79]. In this approach, transform - limited pulses can be realized for any spectral phase signature of coherent combs by appropriately compensating the relative phase of the different comb lines (Figure 16). The ability to achieve successful pulse compression also provides information on the passive stability of the frequency-dependent phase of the Kerr combs. These time domain experiments revealed different pathways to comb formation in terms of phase coherence properties, with the ability to effectively modelock the combs varying significantly depending on many factors such as pump conditions and waveguide dispersion.

Recent studies of the dynamics of comb formation [44] (Figure 17) have shown that initial oscillation often begins at resonances near the parametric gain, or modulational instability, peak with spacings varying widely from the FSR to values as wide as 50nm to 100nm. Cascaded FWM then replicates a comb with this initial spacing.

More complex dynamics arise when these comb lines themselves seed their own mini-comb bands based on the local dispersion and pumping conditions, often with a spacing at or near the cavity FSR in that wavelength region. While these "sub-combs" maintain coherence within themselves, they are not coherently related to the sub-combs generated by other lines of the initial, more widely spaced, comb, and so this results in broadband comb with complex and generally limited coherence. One approach to achieve high coherence is to design the device such that the initial comb intrinsically oscillates at the FSR, producing a wideband comb with high coherence properties. Figure 17 shows the output of a ring resonator under different pumping conditions [44], dramatically showing the transition to from an incoherent state where the frequency sub-combs are well separated and hence incoherent, to a coherent state resulting in a much improved ability to lock the modes to produce optical pulses, resulting in much lower RF noise in the output.

Further studies [49] of the temporal and optical and radio-frequency spectral properties of a parametric frequency combs generated in SiN microresonators demonstrate that the system undergoes a transition to a modelocked state and that ultrashort pulse generation occurs. From a 25-nm filtered section of the 300-nm comb spectrum, sub-200-fs pulses are observed and calculations indicate that the pulse generation process is consistent with soliton modelocking, which is consistent with very recent work involving comb generation in $MgF_2$ microresonators [91] and could be explained by the formation of temporal cavity solitons [91, 92], where contributions from dispersion and loss are compensated by nonlinearity and a coherent driving beam.

## 8. Ultrashort pulsed modelocked lasers

The first demonstration of stable modelocking of a frequency comb based on a microresonator was recently achieved (Figure 18) [15, 50, 93] by embedding the resonator in an active fiber loop. The unique aspect of this system is that the microresonator is used as both a linear filter and as the nonlinear element. This scheme, termed Filter-Driven Four-Wave-Mixing (FD-FWM), is an advancement inspired by dissipative FWM (DFWM) [94 - 96] where the nonlinear interaction occurs in-fibre and is then "filtered" separately by a linear FP filter. This new approach is intrinsically more efficient and so allows for substantially reduced main cavity lengths which in turn has enabled the achievement of highly stable operation – something that has so far eluded DFWM based lasers. A fundamental challenge with DFWM lasers is that the main cavity mode spacing is typically much finer than the microcavity, allowing many cavity modes to oscillate within each micro-resonator mode, giving rise to so-called "supermode instability". The FD-FWM approach allows a substantial reduction in the main cavity length to the point where only a very small number (< 3) cavity modes – and even a single mode - exist within each microresonator resonance, allowing for the possibility of only one main cavity mode oscillating within each ring resonance. Hence, FD-FWM has achieved highly stable operation at high repetition rates over a large range of operating conditions, without a stabilization system, robust to external (i.e., thermal) perturbations. In addition, it can potentially produce much narrower linewidths than ultrashort cavity mode-locked lasers because the long main cavity results in a much smaller Schawlow-Towns phase noise limit [97]. Figure 19 compares the optical spectra, time resolved optical waveforms (measured by autocorrelation) and radio-frequency (RF) spectra of short and long cavity length lasers, clearly showing that the RF spectra for the short cavity laser is highly stable whereas the long cavity length laser not only shows large, long timescale, instabilities in the RF output, but is also unstable on a short timescale where the optical autocorrelation traces show a very limited contrast ratio – a hallmark of unstable oscillation. Recently [50] stable modelocked laser operation with two modes oscillating within each ring cavity was demonstrated, which produced a highly pure RF tone over the modelocked train, enabling high resolution RF spectral measurement of the optical linewidth, which achieved the remarkably narrow width of about 10kHz.

## 9. Ultrafast Phase Sensitive Pulse Measurement

Coherent optical communications [18, 19] have created a compelling need for ultrafast phase-sensitive measurement techniques operating at milliwatt peak power levels. Ultrafast optical signal measurements have been achieved using time-lens temporal imaging on a silicon chip [98, 99] and waveguide-based Frequency-Resolved Optical Gating (FROG) [106], but these are either phase insensitive or require waveguide-based tunable delay lines - a difficult challenge. Recently [14], a device capable of both amplitude and phase characterization of ultrafast optical pulses was reported, operating via a novel variation of Spectral Phase Interferometry for Direct Electric-Field Reconstruction (SPIDER) [101 - 109] based on FWM in a Hydex waveguides. Here, pulse reconstruction was obtained with the aid of a synchronized incoherently related clock pulse.

SPIDER approaches are ultrafast, single-shot, and have a simple, direct and robust phase retrieval procedure. However they traditionally have employed either three-wave mixing (TWM) or linear electro-optic modulation, both of which require a second-order nonlinearity that is absent in CMOS compatible platforms. Also, typical SPIDER methods work best with optical pulses shorter than 100 fs (small time-bandwidth products (TBP)) and peak powers > 10 kW, and hence are not ideally suited to telecommunications. The device reported in [14] measured pulses with peak powers below 100 mW, with a frequency bandwidth greater than a terahertz and on pulses as wide as 100 ps, yielding a record TBP > 100 [106].

Figure 20 shows a schematic of the device. The pulse under test (PUT) is split into two replicas and nonlinearly mixed with a highly chirped pump inside the chip. The resulting output is then captured with a spectrometer and numerically processed to extract the complete information (amplitude and phase) of the incident pulse. Figure 21 compares the results from the X-SPIDER device using both a standard algorithm and a new extended (Fresnel) phase-recovery extraction algorithm [14, 109] designed for pulses with large TBP, with results from SHG-based FROG measurements. As expected, for low TBP pulses (short-pulse regime) the SPIDER device yielded identical results with both algorithms and agreed with the FROG spectrogram. For large TBP pulses (highly chirped, long pulsewidths) the X-SPIDER results obtained with the new algorithm agreed very well with the FROG trace, whereas results using the standard phase-recovery algorithm were unable to accurately reproduce the pulse.

## 10. Conclusions

We have reviewed the recent progress in CMOS-compatible, non c-Si integrated optical platforms for nonlinear optics, focusing on SiN, Hydex glass, and amorphous Si.

These new platforms have enabled the realization of at telecom wavelength bands. The combination negligible nonlinear (two-photon) absorption, low linear loss, the ability to engineer dispersion, and moderately high nonlinearities has enabled these new platforms to achieve new capabilities and novel devices not possible in c-Si because of its low FOM. These include on-chip multiple wavelength oscillators, optical frequency comb sources, ultrashort optical pulse generators, high-gain optical parametric amplifiers, phase-sensitive optical pulse measurement and many others. These platforms will likely have a significant impact on future all-optical devices, complementing the capabilities already offered by silicon nanophotonics. The high performance, reliability, manufacturability of all of these platforms combined with their intrinsic compatibility with electronic-chip manufacturing (CMOS) raises the prospect of practical platforms for future low-cost nonlinear all-optical PICs that offer a smaller footprint, lower energy consumption and lower cost than present solutions.

## Table I

Nonlinear parameters for CMOS compatible optical platforms

|  | a-Si [126] | c-Si [2, 47] | SiN [55-57] | Hydex [63, 76] |
|---|---|---|---|---|
| $n_2$ (x fused silica[1]) | 700 | 175 | 10 | 5 |
| $\gamma$ [W$^{-1}$m$^{-1}$] | 1200 | 300 | 1.4 | 0.25 |
| $\beta_{TPA}$ [cm/GW] | 0.25 | 0.9 | negligible[2] | negligible[3] |
| FOM | 5 | 0.3 | >> 1 | >> 1 |

[1] $n_2$ for fused silica = 2.6 x 10$^{-20}$ m$^2$/W [1]

[2] no nonlinear absorption has been observed in SiN nanowires.

[3] no nonlinear absorption has been observed in Hydex waveguides up to intensities of 25GW/cm$^2$ [63].

## Table II

| Reference | [104] | [102,103] | [99] | [98] | [100] | [101] |
|---|---|---|---|---|---|---|
| $n_2$.[10$^{-17}$ m$^2$/W] | 2.1 | 1.3 | 4.2 | 0.05 | 0.3 | 7.43 |
| $\gamma$ [W$^{-1}$m$^{-1}$] | 1200 | 770 | 2000 | 35 | N/A | N/A |
| $\beta_{TPA}$ [cm/GW] | 0.25 | 0.392 | 4.1 | 0.08 | 0.2 | 4 from [96,98] |
| FOM | 5±0.3 | 2.2±0.4 | 0.66±0.3 | 0.4 | 0.97 | 1.1 |

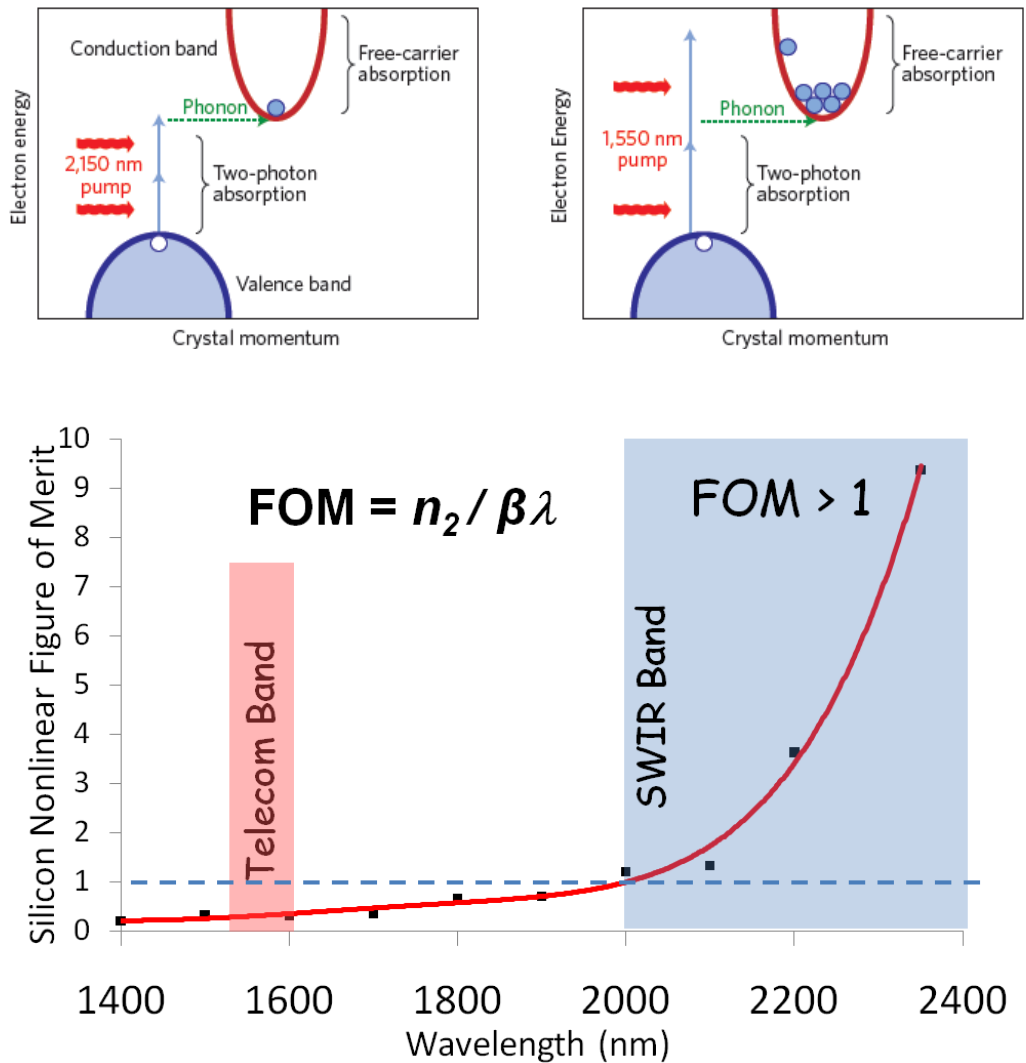

**Figure 1.** Top: source of two-photon absorption in silicon in the telecom band near 1550nm via indirect transitions. Bottom: nonlinear FOM of silicon from 1400nm to 2400nm obrained by taking an average of results reported in the literature [25-27], showing that the FOM is well below 1 in the telecom band due to the indirect TPA, and exceeds 1 beyond 2000nm, making it an attractive nonlinear platform in this wavelength range .

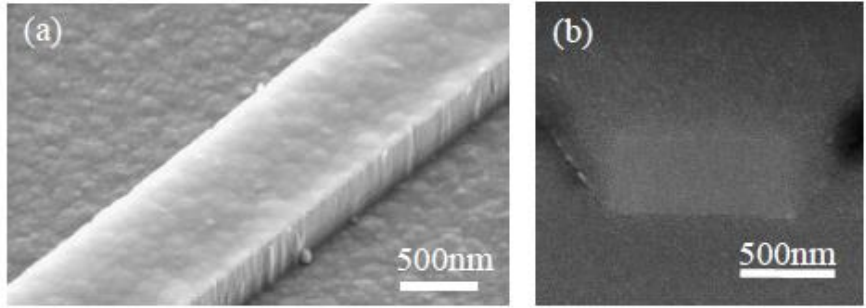

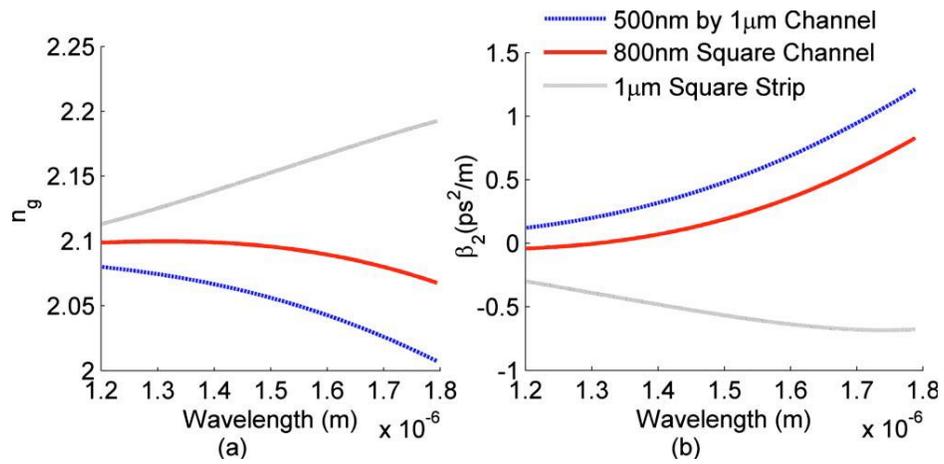

**Figure 2.** Silicon nitride nanowires (a) SEM micrograph of the SiN/SiO2 waveguides before and after the SiO2 upper-cladding deposition, respectively [34]; Bottom: calculated group index (b) and group velocity dispersion (c) of SiN nanowires with different geometries from [35].

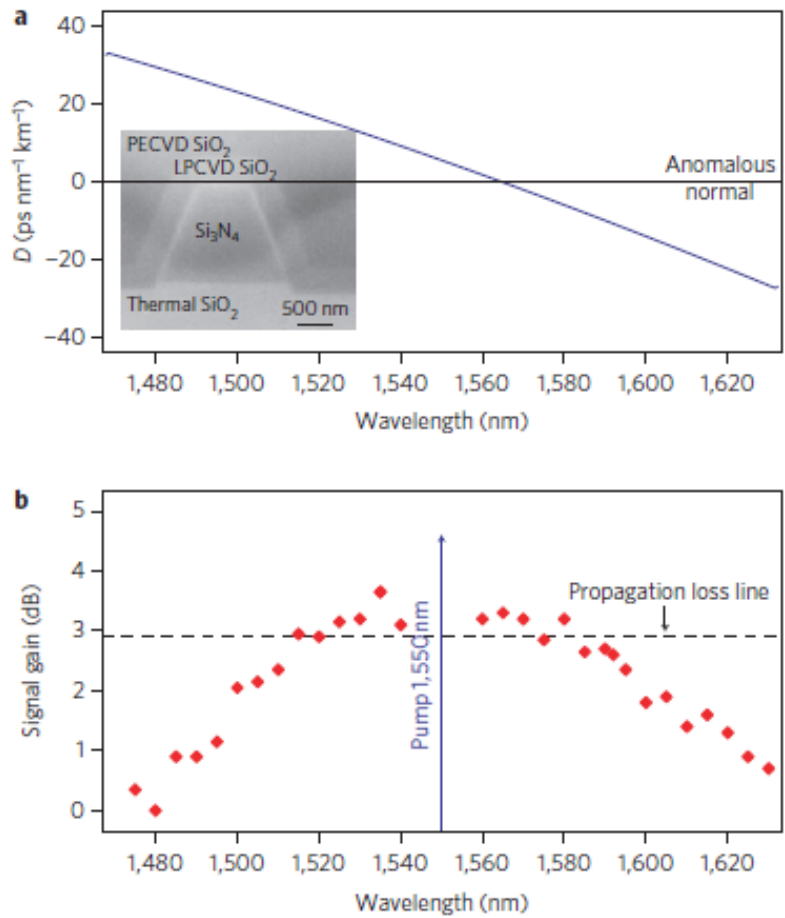

**Figure 3.** Silicon nitride nanowires. Top: Theoretical dispersion [36] of a SiN nanowire grown by low pressure CVD, showing a zero-GVD point at 1,560 nm with anomalous dispersion in the C-band. Inset shows a scanning electron micrograph (SEM) of the cross-section of the SiN nanowire highlighting the trapezoidal shape of the core and the cladding materials; Bottom: parametric gain achieved with a pump wavelength at 1550nm, from [36].

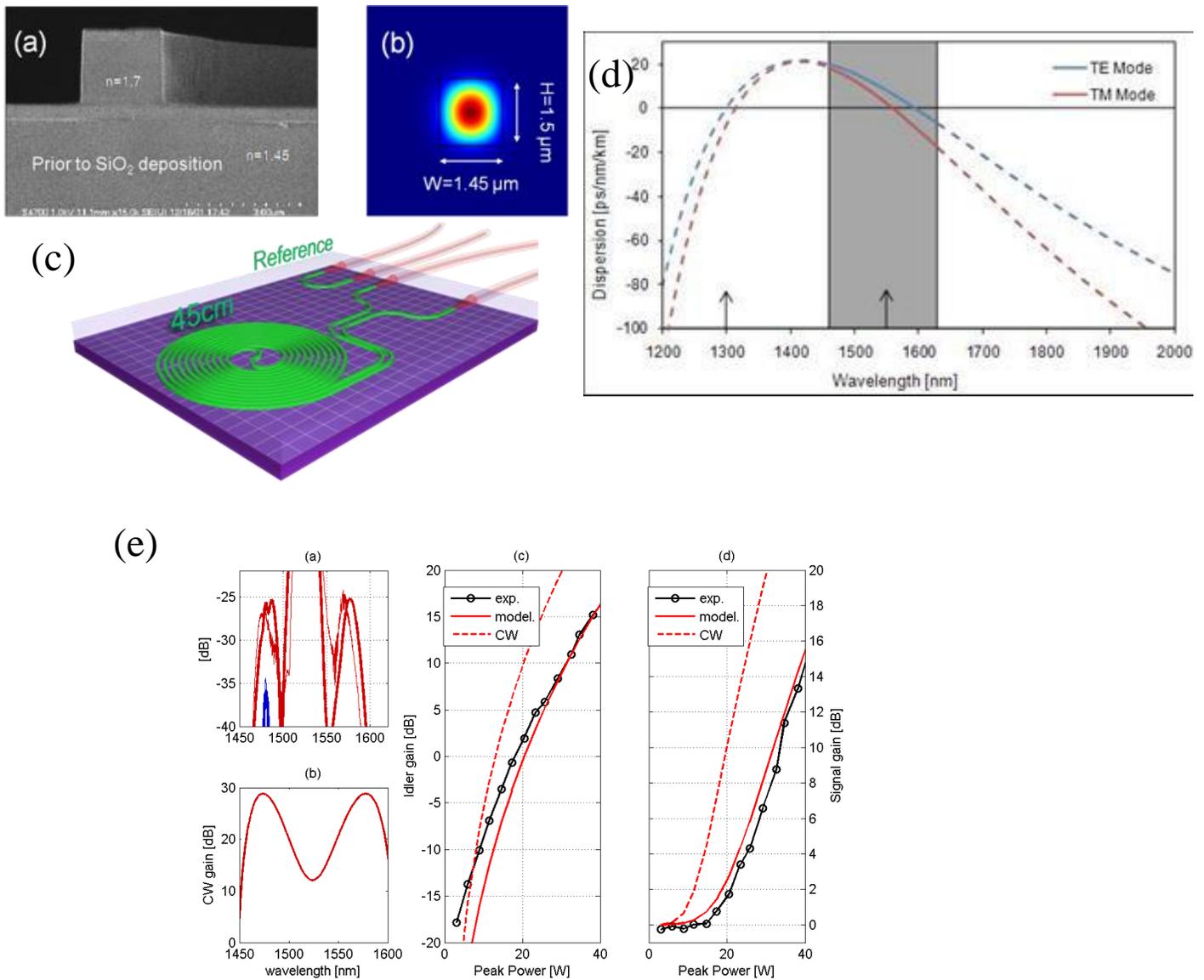

**Figure 4.** Hydex waveguides. (a) SEM image of the cross-section of the Hydex waveguide [14] prior to the final deposition of the SiO2 upper cladding; (b) Theoretical mode profile and (c) top down schematic view of the 45cm long spiral waveguide; (h) dispersion curves for the quasi-TE and quasi-TM modes of the Hydex waveguides [54]. The shaded region indicates the experimentally measured wavelength range (solid lines); the dispersion is extrapolated outside this region (dashed lines). The two zero dispersion points are illustrated by the vertical arrows (e) parametric gain curves from [53] when pumping with a pulsed source at peak powers up to 40W with a wavelength at 1540nm.

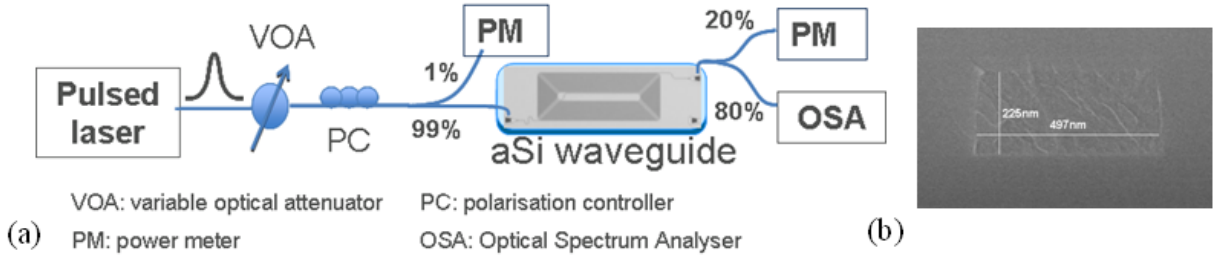

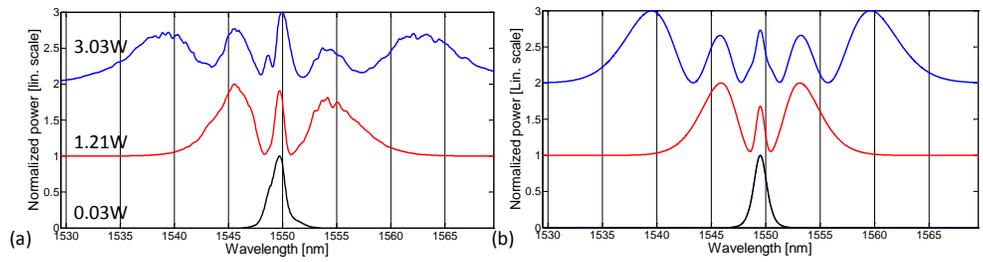

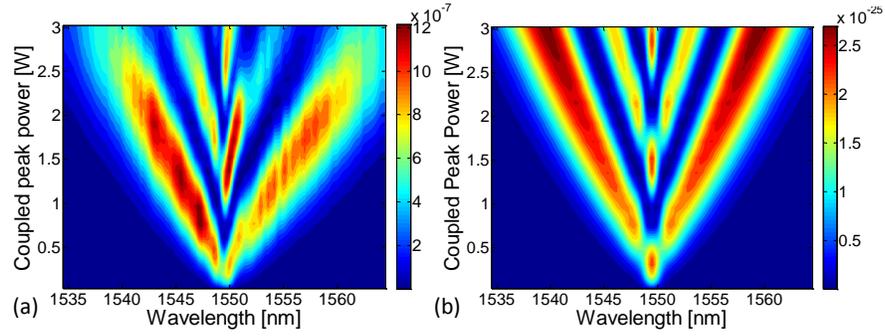

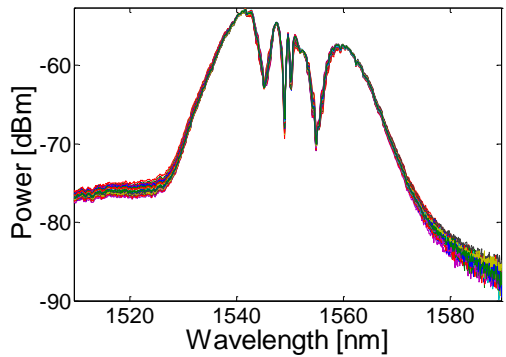
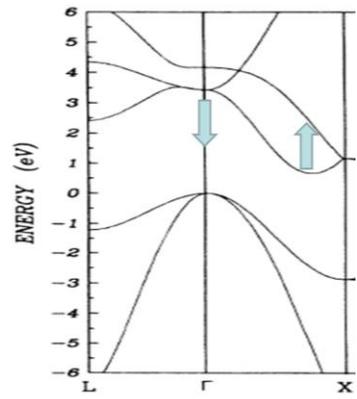

**Figure 5.** Amorphous silicon nanowires from [65]. Top: Experimental setup for measuring self phase modulation and nonlinear transmission in order to extract the Kerr nonlinearity ($n_2$) and the two-photon absorption coefficient. Output spectra for 0.03W, 1.21W and 3.03W coupled peak power - (a): Experiment, (b): Simulation. The curves are normalized and shifted upwards with increasing powers for clarity. (a) Experimental and (b) theory 2D plots showing the spectral broadening of the output pulse spectra vs coupled peak power. Note the linear intensity scale at the right is relative. Bottom: self

phase modulation broadened spectrum taken over a period of an hour, showing negligible change.

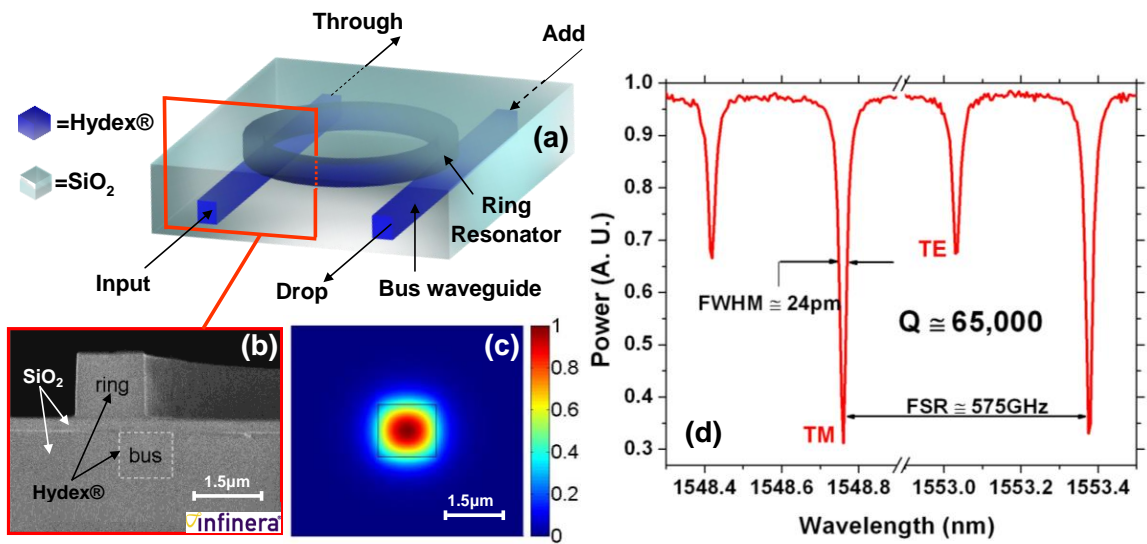

**Figure 6.** Hydex micro-ring resonator [38] with moderate Q factor = 65,000, resulting in a bandwidth of 3GHz, with an FSR of 575GHz.

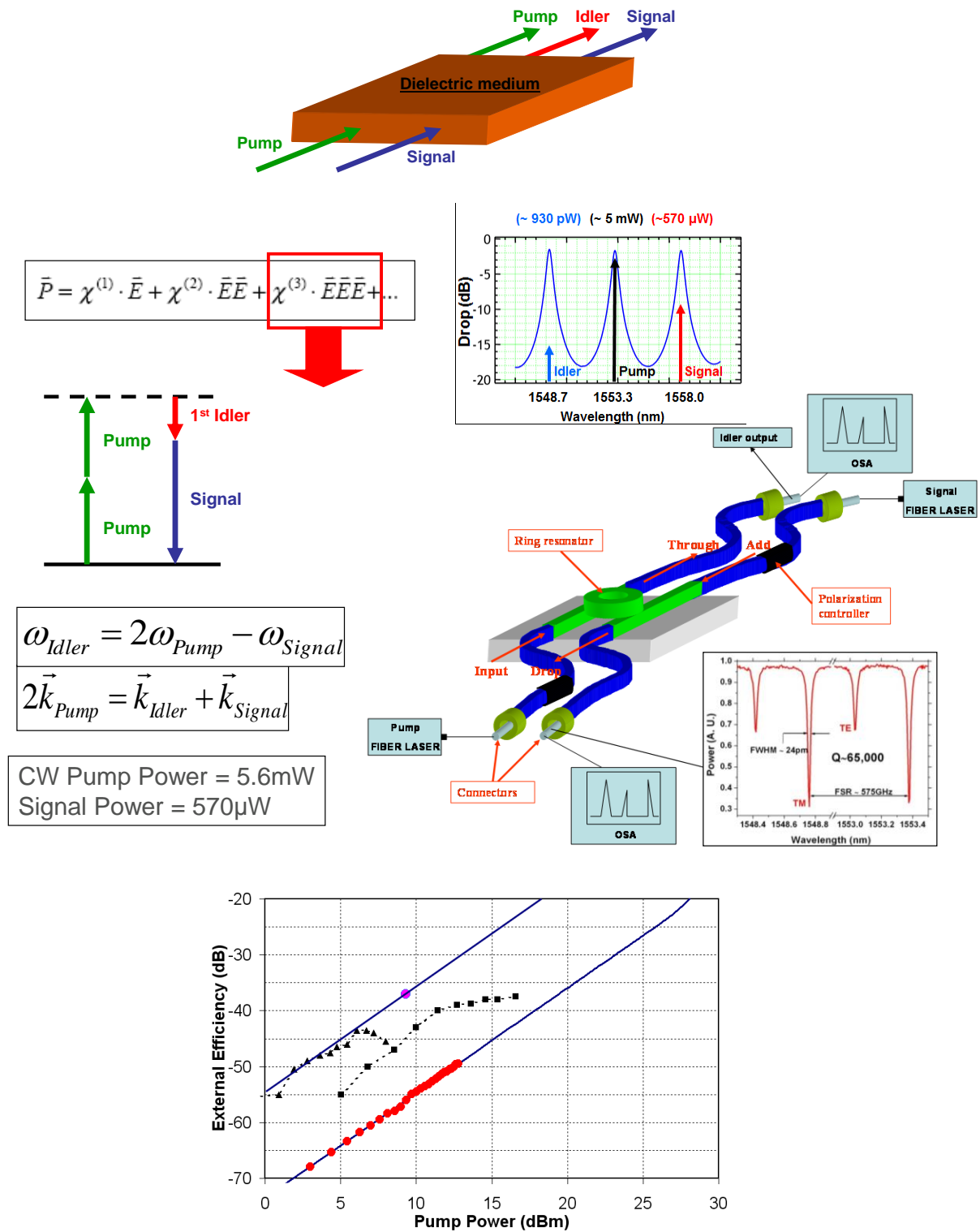

**Figure 7.** Low power FWM in Hydex ring resonators with moderate Q factor = 65,000, resulting in a bandwidth of 3GHz, with an FSR of 575GHz [38]. Bottom: External conversion efficiency along with estimated efficiency of silicon RR reported in [72].

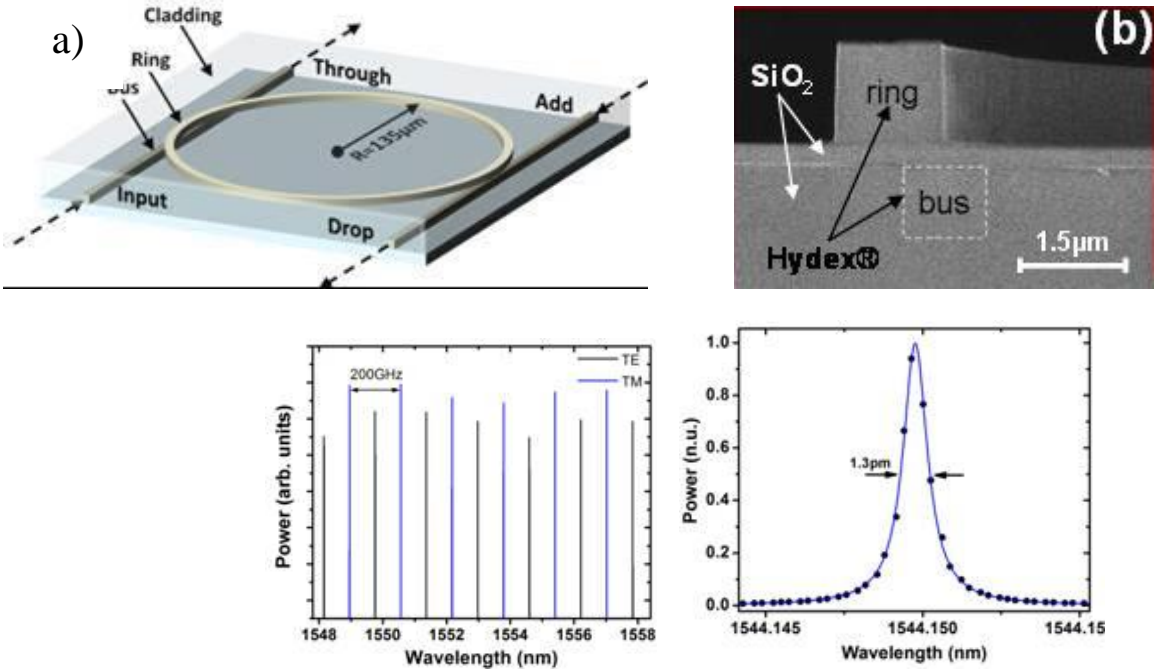

**Figure 8.** Integrated ring resonators based in Hydex [37]. (a) Hydex four-port microring resonator (fibre pigtails not shown) with Q=1.2x10$^6$ . (b) SEM image of the Hydex ring resonator cross section before depositing the upper cladding of SiO2; (c) Linear transmission through ring resonator from the INPUT to THROUGH port for TM and TE polarizations.

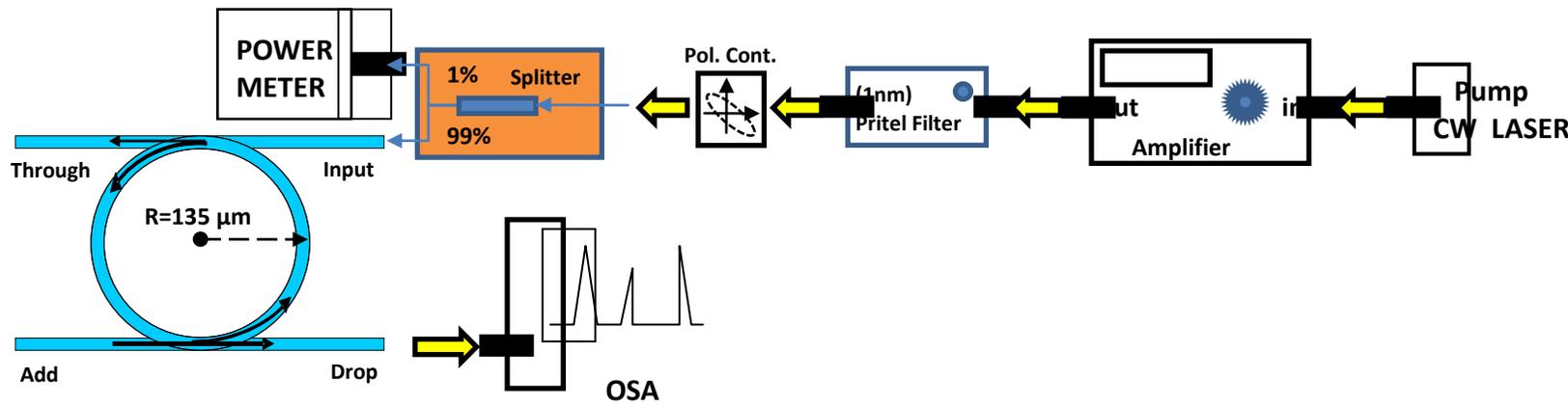

**Figure 9.** Experimental setup for OPO via external pumping with soft thermal locking [37].

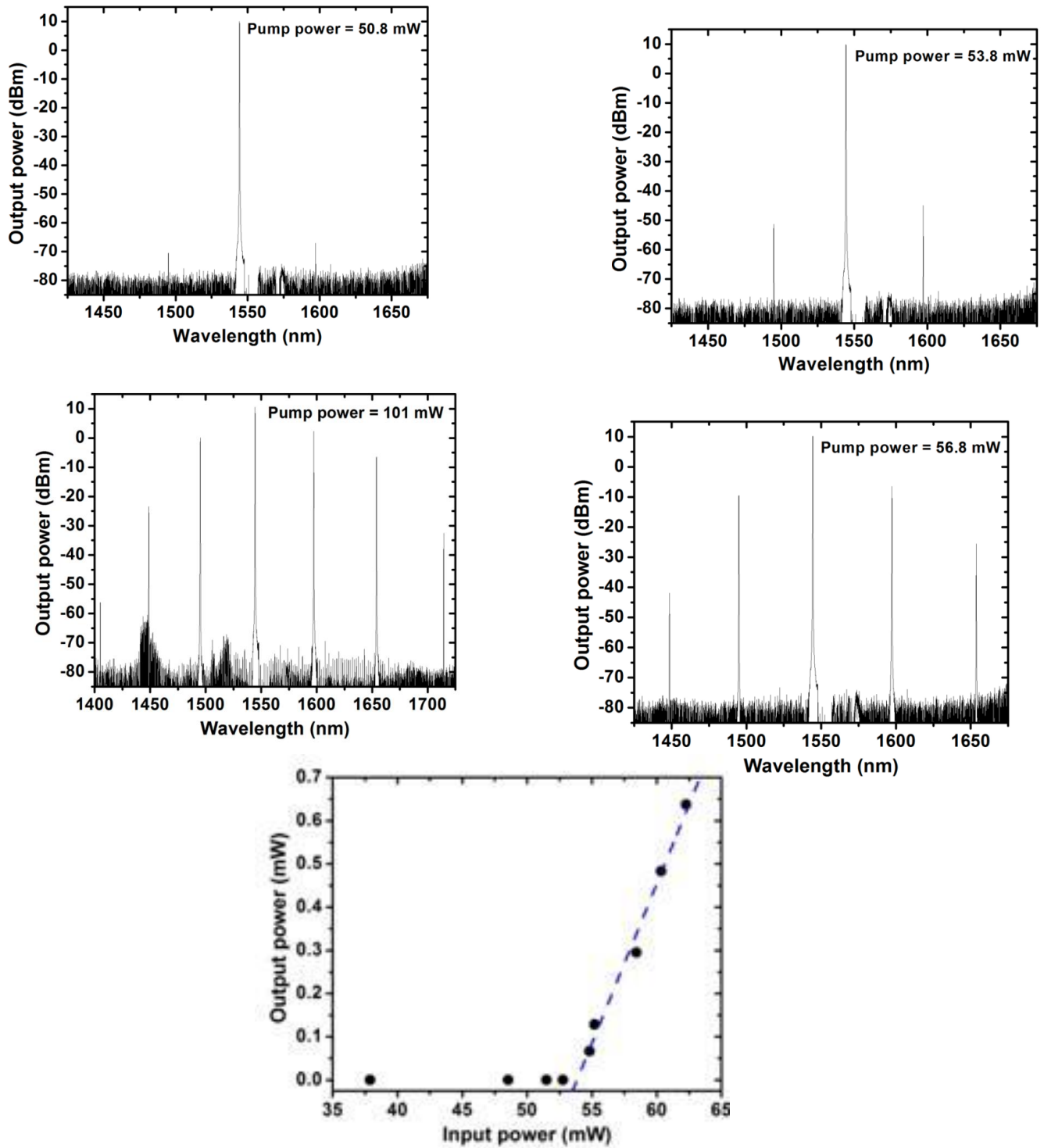

**Figure 10.** Output spectra of Hydex hyper-parametric oscillator [37] with a pump power at 50.8mW (just above threshold) tuned to a resonance at 1,544.15 nm (TM poloarization) (d) same but with a pump power well above threshold at 101 mW (f) Output power vs pump power (at 1,544.15 nm) in the drop port waveguide of the Hydex oscillator for a single oscillating mode at 1,596.98 nm, showing a threshold of 54mW with a differential slope efficiency of 7.4% (linear fit - blue dashed line) above threshold.

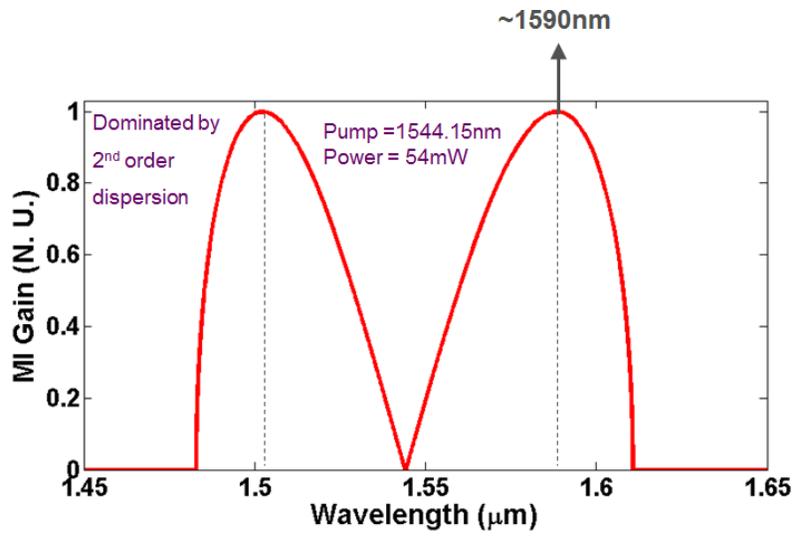

**Figure 11.** Theoretical calculation of modulational instability (MI) gain [37] based on the measured dispersion, at a pump power of 54mW and pump wavelength near the zero dispersion point at 1544nm, showing a gain peak about 46nm away from the pump wavelength

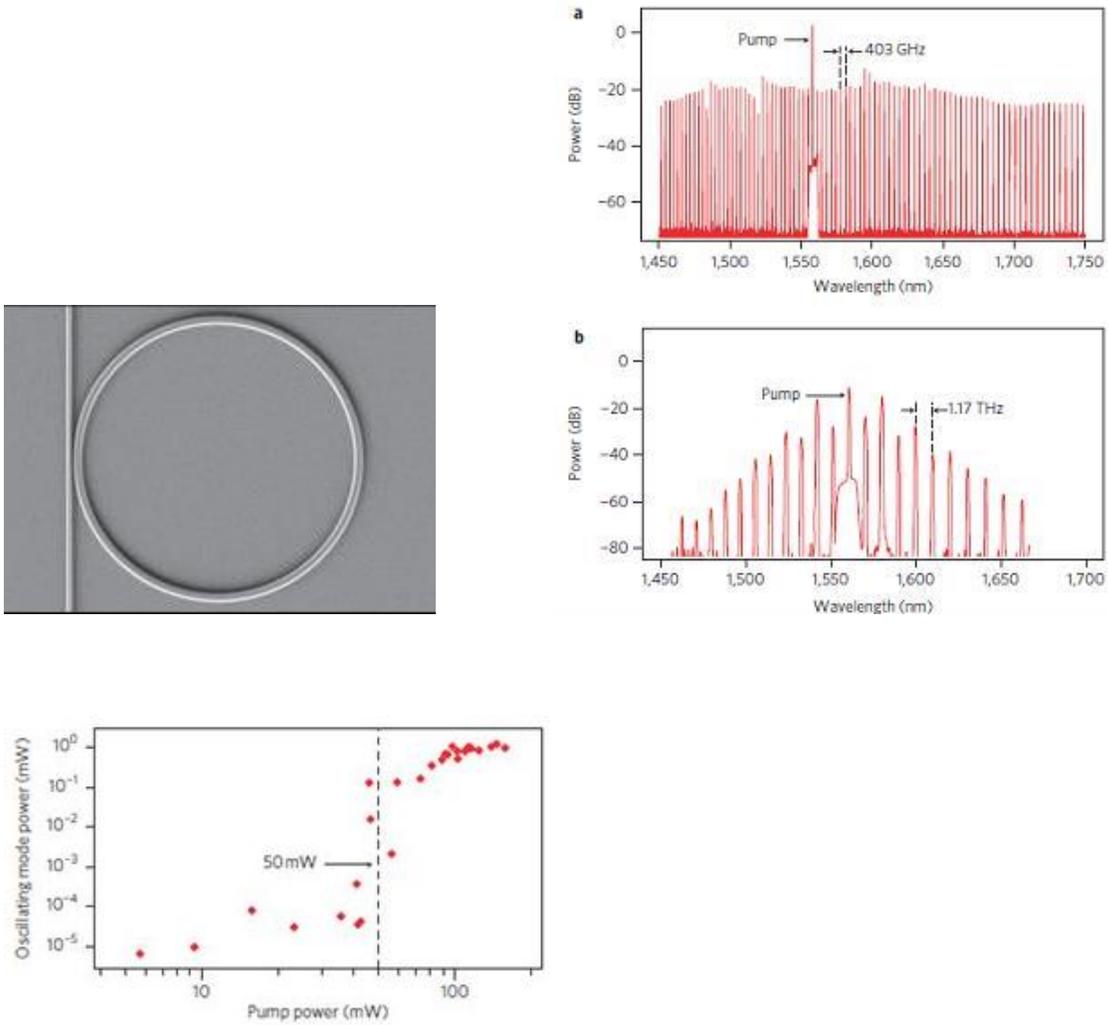

**Figure 12. Integrated OPO multiple wavelength sources in SiN [36] (a-c) ring resonators [36].** SEM image of SiN microring resonator (58μm radius and Q= 500,000, FSR= 403 GHz) from [36] coupled to a bus waveguide, with cross section height of 711 nm, base width of 1,700 nm and a sidewall angle of 20 degrees, giving anomalous GVD in the C-band and a zero-GVD point at 1,610 nm; (a) Output spectra of a 58-μm-radius SiN ring resonator OPO with a single pump wavelength tuned to a resonance at 1,557.8 nm, numerous narrow linewidths at precisely defined wavelengths. The 87 generated wavelengths were equally spaced in frequency, with an FSR of 3.2 nm (b) spectra of a 20-μm radius SiN ring resonator generating a comb with a different frequency spacing, pumped at 1,561 nm and producing 21 wavelengths over a 200-nm span, with a spacing of 9.5 nm (c) output power in the first generated mode of SiN OPO versus pump power, showing a threshold of 50 mW with a slope efficiency of 2%.

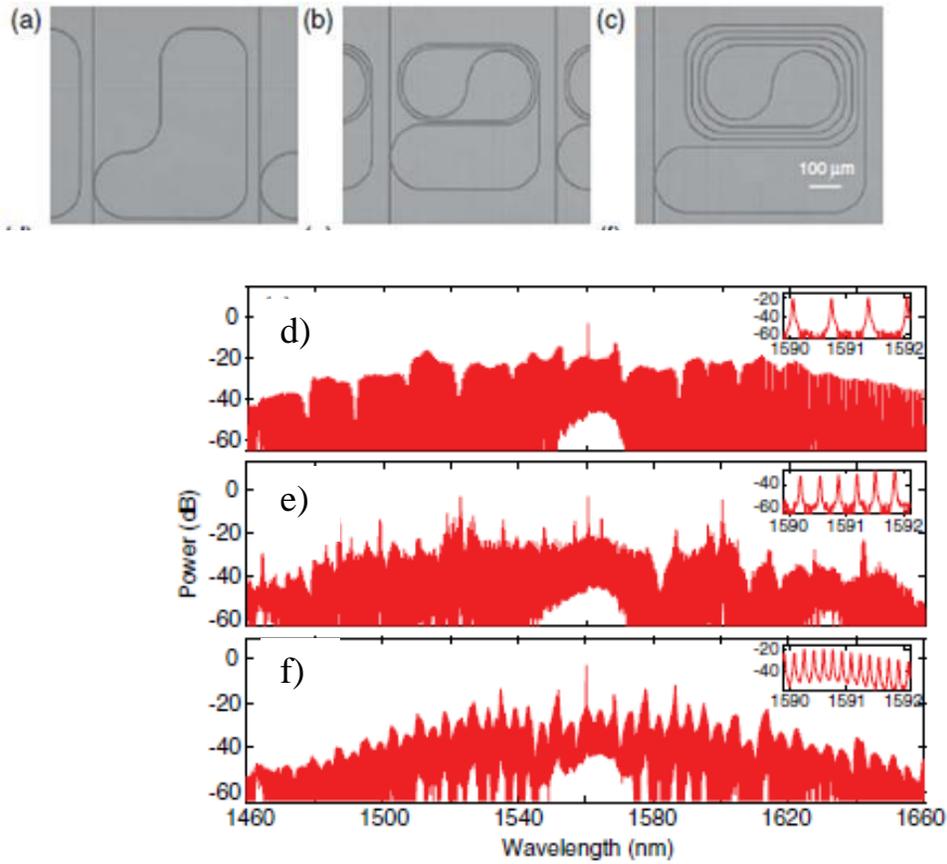

**Figure 13. Advanced frequency combs in SiN ring resonators [46].** a-f) Sub-100GHz spacing SiN ring resonators. Micrographs of the (a) 80GHz, (b) 40GHz, and (c) 20 GHz FSR resonators and the corresponding linear transmission spectra (d-f). The nanowire cross section was 725nm by 1650 nm, with a microring diameter of 200 µm with a loaded Q of 100,000. Output spectra are 300 nm wide for the 80GHz and 40GHz FSR rings and 200 nm for the 20GHz FSR ring. A 2 nm section of each comb is inset in each figure to illustrate the spacing of the comb lines.

g)

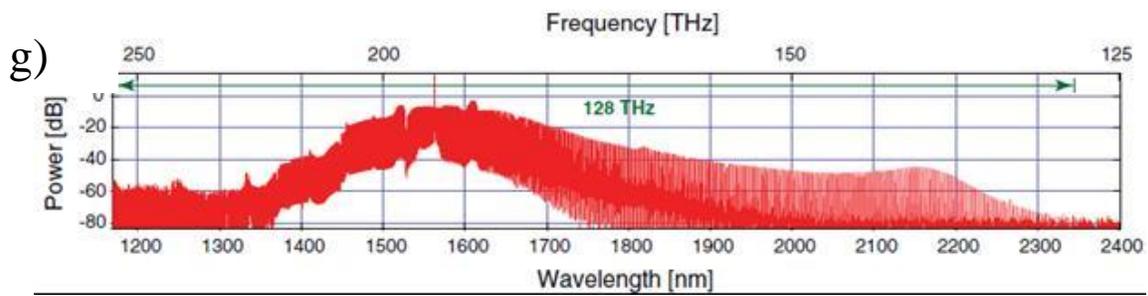

h)

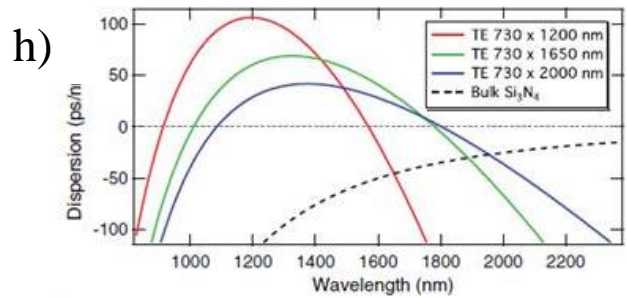

**Figure 14. Advanced frequency combs in SiN ring resonators [45].** Optical spectrum of octave-spanning parametric frequency comb generated in a SiN ring resonator (h) dispersion simulations for the fundamental TE mode of SiN waveguide with a height of 730nm and widths of 1200, 1650, and 2000 nm. The dashed curve shows the dispersion for bulk silicon nitride.

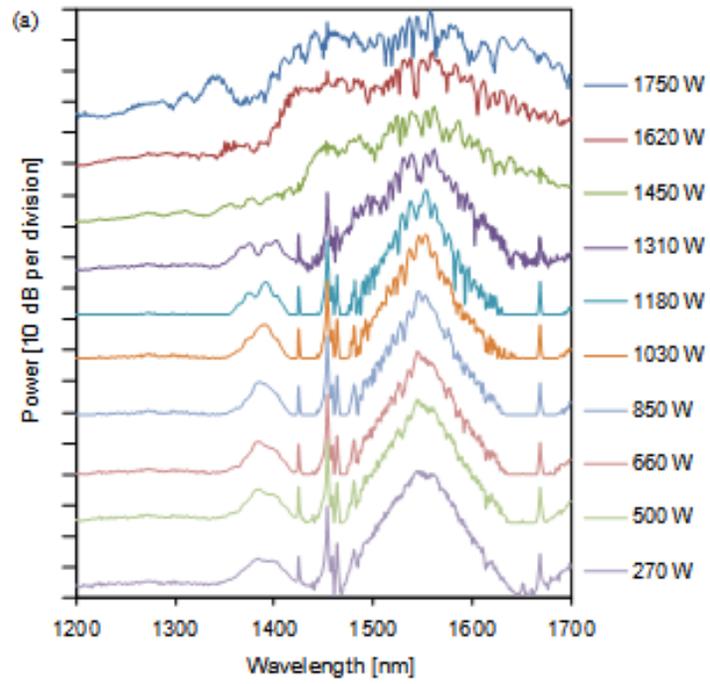

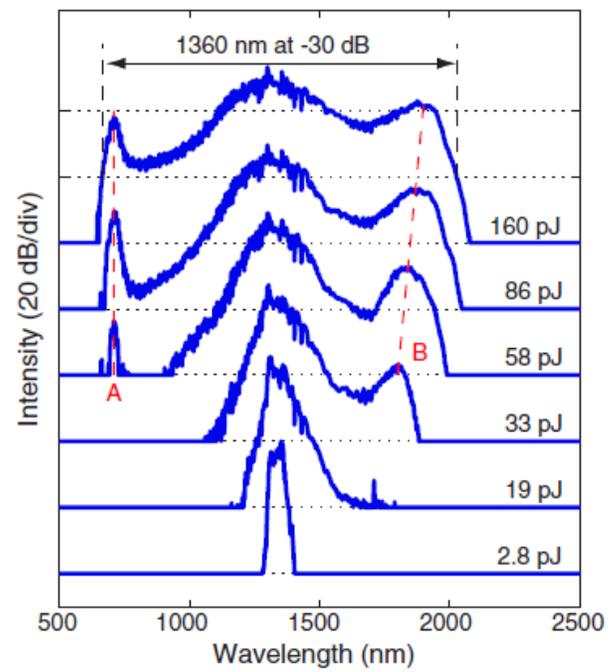

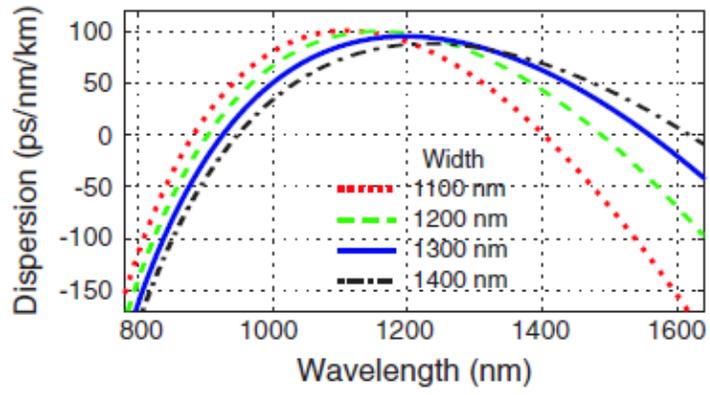

**Figure 15. Supercontinuum Generation.**

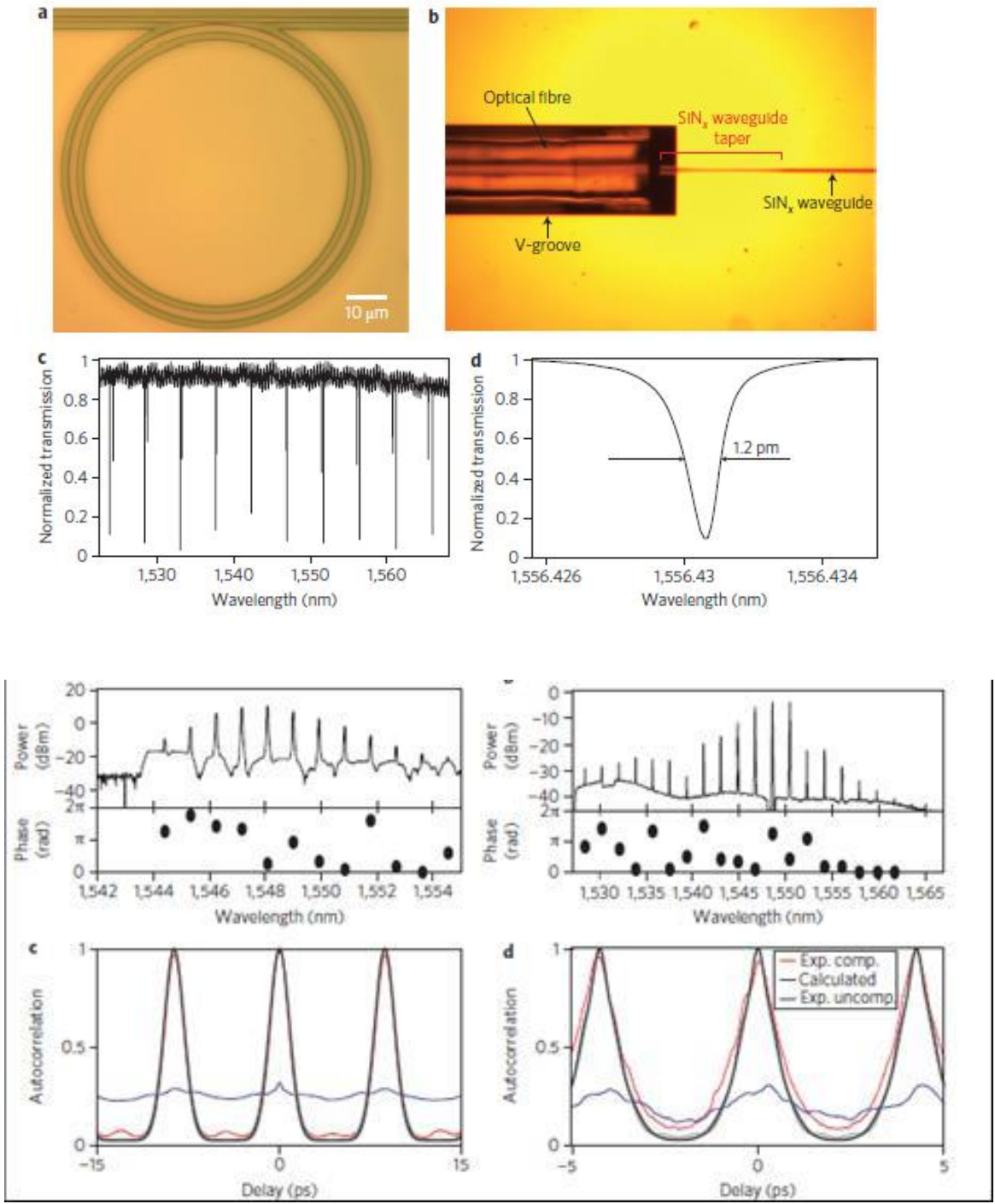

**Figure 16.** g) SEM image of SiN microring resonator from [47] (radius 40 µm) along with (h) fibre-pigtailed coupling waveguide to resonator; (i,j) Optical transmission spectrum for TM polarized light of ring resonator [46] including (j) high resolution spectrum for a mode at 1,556.43 nm showing a linewidth of 1.2 pm, corresponding to a loaded Q factor of $1.3 \times 10^6$ with a FSR of 4.8 nm.

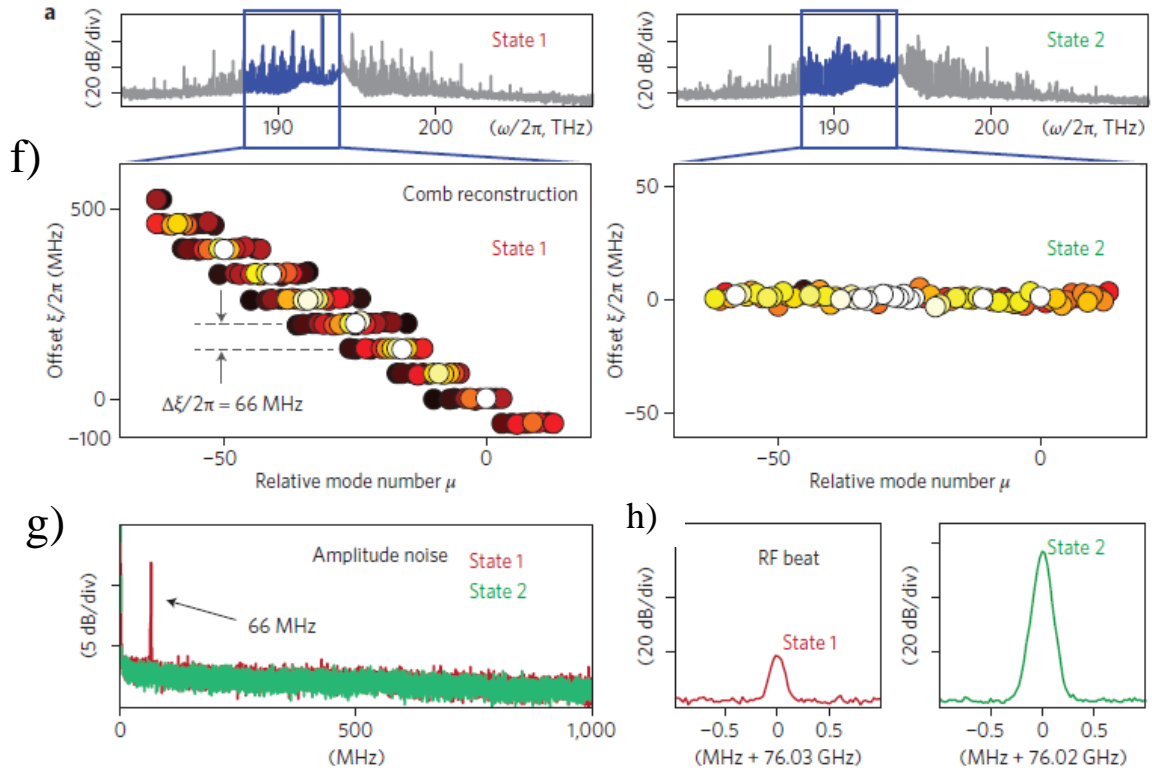

**Figure 17. Coherence, dynamics of frequency comb formation, and ultrashort pulse modelocked sources.** a-d. Study of Frequency comb coherence properties from [48]. a) Output spectra of high-Q silicon nitride microring (Figure 2) showing the ability to compress pulses from type-I Kerr combs. a,b, Spectra of the combs after processing by a pulse shaper, together with the phase applied to the liquid-crystal modulator pixels for optimum SHG signals (c,d) Autocorrelation traces corresponding to (a,b). Red lines are compressed pulses after phase correction, blue lines are uncompressed pulses and black lines are calculated by taking the spectra shown in a and b and assuming a flat spectral phase. The contrast ratios of the autocorrelations measured after phase compensation are 14:1 and 12:1, respectively. Light grey traces show the range of simulated autocorrelation traces. (e-h) Studies of coherence evolution in SiN microring resonators from [43] showing the transition to a low phase noise Kerr comb. e, Optical spectra of two microresonator comb states 1 and 2 (pump power 6W). State 2 evolves from state 1 when reducing the detuning of the pump laser. f, A transition is observed from multiple sub-combs to a single (sub) comb over the bandwidth of the Kerr comb reconstruction. In state 1, all sub-combs have the same mode spacing, but have different offsets which differ by a constant of 66 MHz. g, In the transmission from state 1 to state 2, the amplitude noise peak resulting from the beating between overlapping offset sub-combs disappears

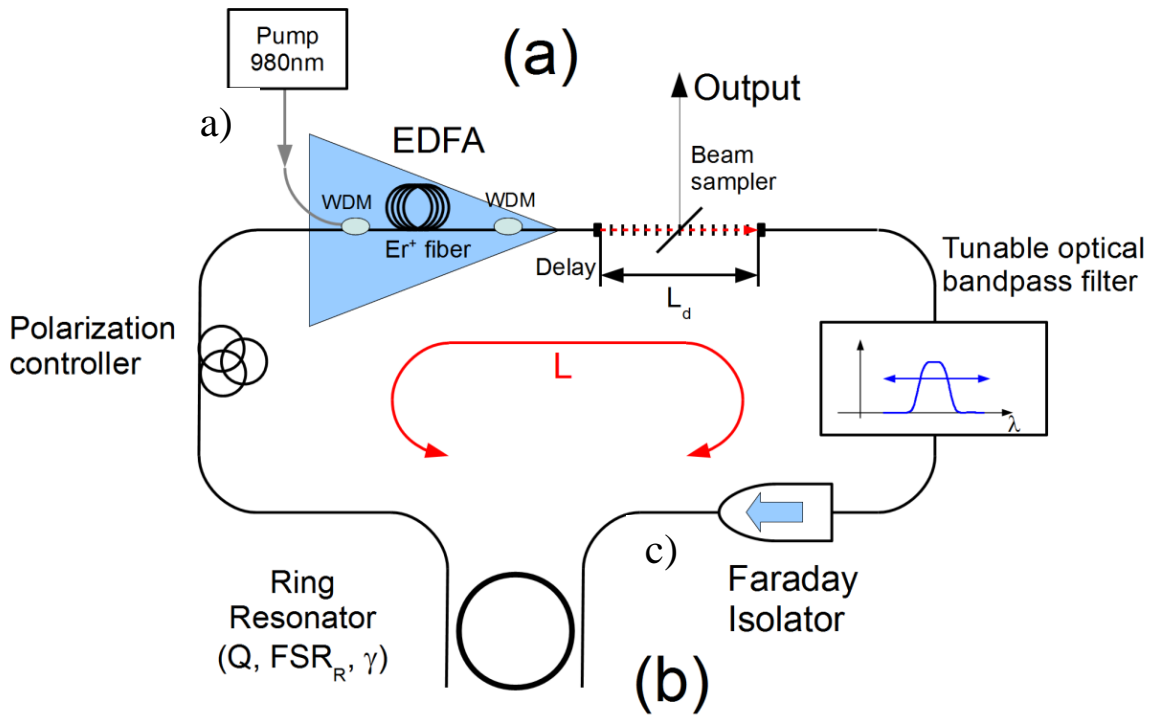

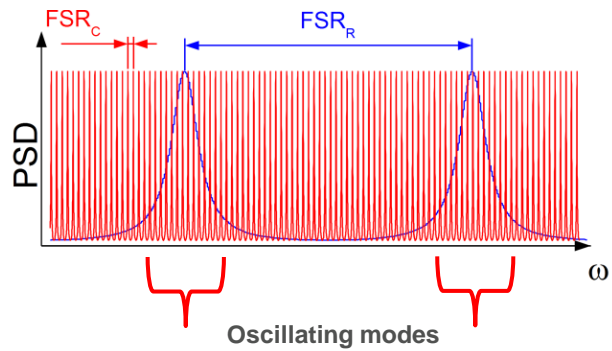

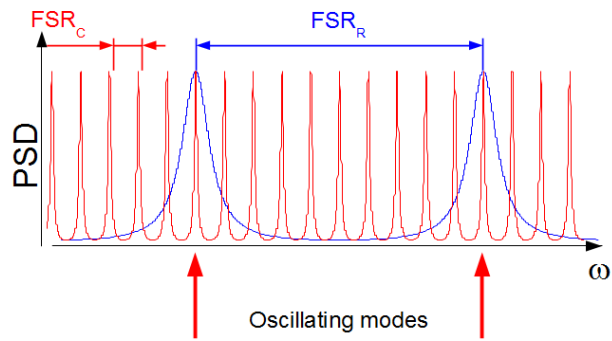

**Figure 18. Ultrafast modelocked laser** based on filter driven four-wave mixing in a modelocked fiber loop laser [15]. (a) Experimental configuration for fiber loop

modelocked laser based on filter driven four wave mixing (FDFWM) based on a microring resonator, where the resonator performs the dual function of linear filtering and nonlinear interaction.

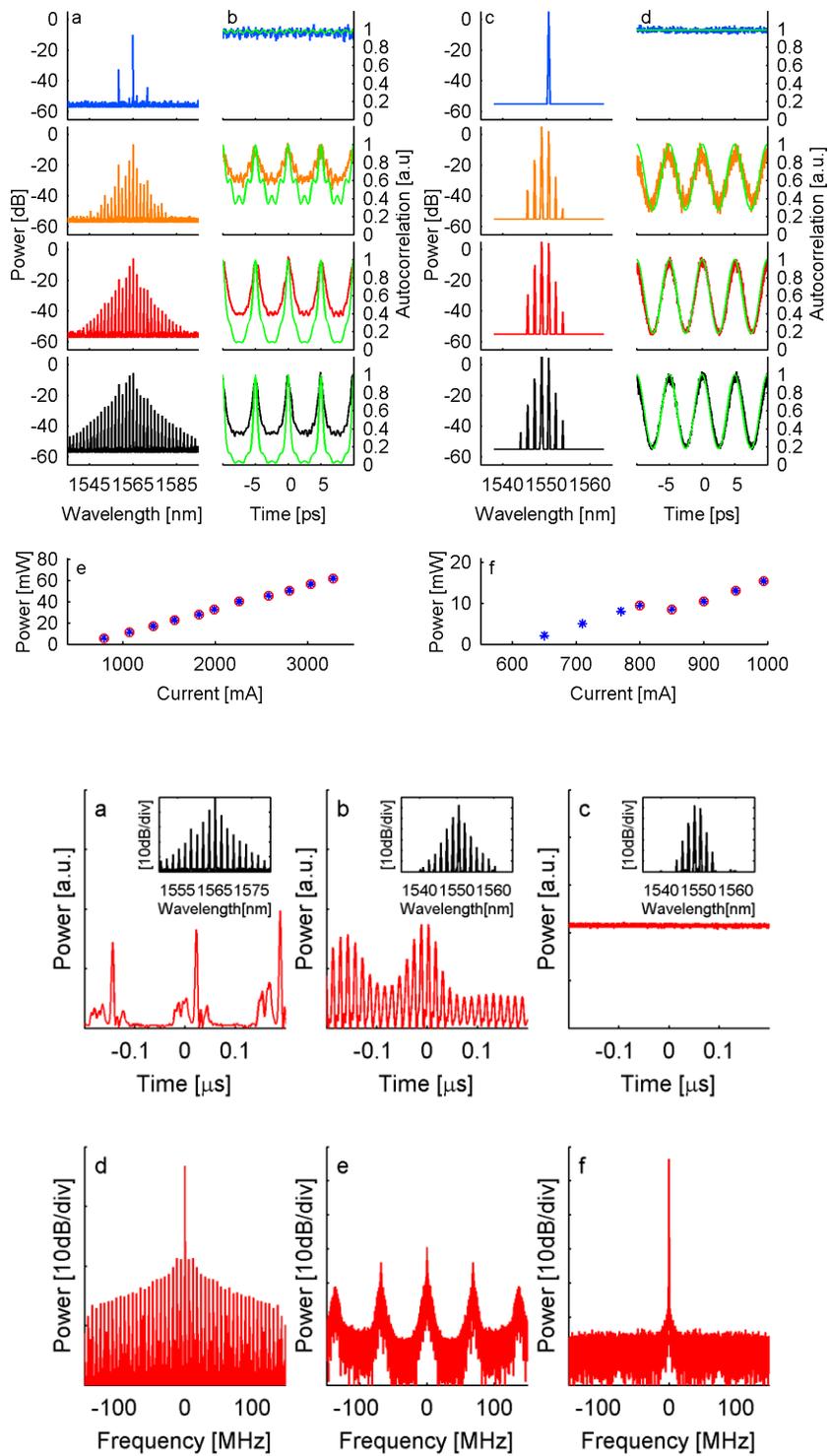

**Figure 19. Ultrafast modelocked laser** based on filter driven four-wave mixing in a modelocked fiber loop laser [15]. Experimental optical temporal waveforms measured

via optical autocorrelator and radio frequency (RF) temporal (d,e) and RF spectral (f,g) output of the fiber loop laser for two different main cavity lengths. (b, d, f) are for long cavity (L=33m, FSR = 6.0 MHz) laser and (c, e, g ) are for short cavity main length (L=3m, FSR=68.5MHz) laser. The autocorrelation plots show theoretical calculations (green) starting from the experimental optical spectra for a fully coherent and transform-limited system calculated by considering each line of the experimental optical spectra as being perfectly monochromatic and in-phase with the others, yielding an output pulse with a FWHM of 730 fs for the highest excitation power condition. The measured autocorrelation for the long cavity laser (b) shows a considerably higher background than the expected autocorrelation (50:1). Conversely, the calculated autocorrelation trace for the short-cavity laser (c) perfectly matches the measured trace, indicating stable modelocking, and corresponding to a transform-limited pulse width (FWHM) of 2.3 ps with a peak to background ratio of 5:1. Stable oscillation was obtained by adjusting the phase of the cavity modes for the short main cavity length laser relative to the ring resonator modes via a free space delay line.

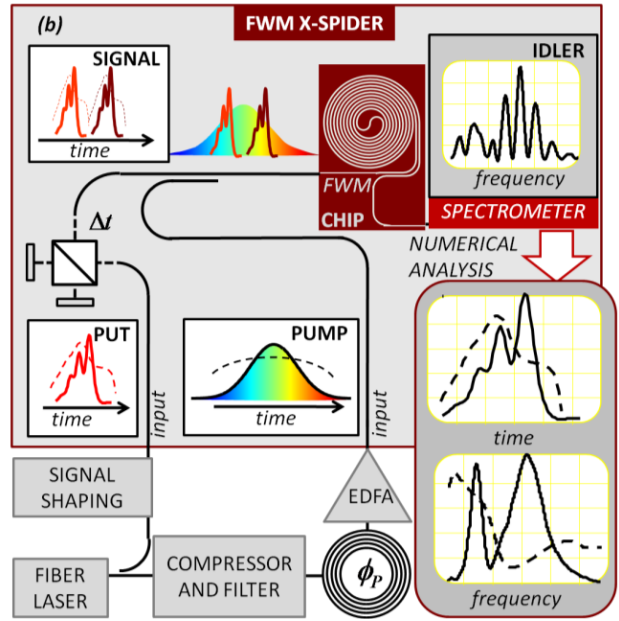

**Figure 20. Experimental setup for the phase measurement of Ultrafast Pulses.** (h, i) X-SPIDER device based on FWM in Hydex spiral waveguide [14].

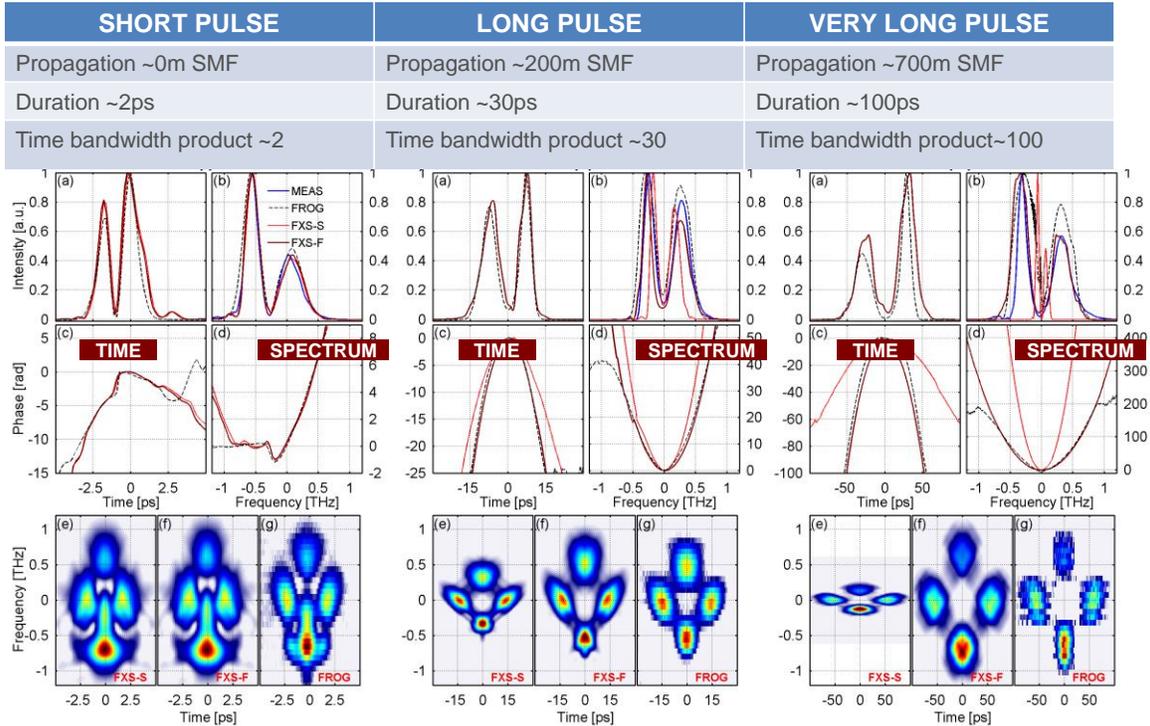

| SHORT PULSE | LONG PULSE | VERY LONG PULSE |
|---|---|---|
| Propagation ~0m SMF | Propagation ~200m SMF | Propagation ~700m SMF |
| Duration ~2ps | Duration ~30ps | Duration ~100ps |
| Time bandwidth product ~2 | Time bandwidth product ~30 | Time bandwidth product~100 |

**Figure 21. Measurement of Ultrafast Pulses.** X-SPIDER device based on FWM in Hydex spiral waveguide [14]. Retrieved phase and amplitude profiles for pulses with time bandwidth products (TBP) of 5 (h) and 100 (i). The left hand plots were obtained by the X-SPIDER device using a standard algorithm while the middle plots used a new algorithm [14] for large time-bandwidth product (TBP) pulses. The plots at the right are experimentally measured FROG SHG measurements.